\documentclass[iop]{emulateapj}

\def\LIR{\hbox{$L_{\rm IR}$}}

\def\HII{\hbox{H\,{\sc ii}}}

\def\NeII{\hbox{[Ne\,{\sc ii}]12.81\,\micron}}

\def\H2{\hbox{H$_{2}$}}
\def\PAHa{\hbox{11.3\,\micron\,PAH}}

\def\PAHd{\hbox{6.2\,\micron\,PAH}}

\def\Halpha{\hbox{\rm H$\alpha$}}

\def\deg{$^{\circ}$}

\def\Lsun{\hbox{$L_\odot$}}
\def\Msun{\hbox{$M_{\odot}$}}
\def\LIR{\hbox{$L_{\rm IR}$}}

\def\MH2{\hbox{$M_{H_2}$}}

\def\mum{\hbox{$\mu$m\ }}

\shorttitle{The Extended mid-Infrared Emission in (U)LIRGs}
\shortauthors{D\'{\i}az-Santos et al.}

\begin{document}

\title{The Spatial Extent of (U)LIRG\lowercase{s} in the mid-Infrared I: The Continuum Emission}



\author{T.~D\'{\i}az-Santos\altaffilmark{1},
V.~Charmandaris\altaffilmark{1,2},
L.~Armus\altaffilmark{3},
A.~O. Petric\altaffilmark{3},
J.~H.~Howell\altaffilmark{3},
E.~J.~Murphy\altaffilmark{3},
J.~M.~Mazzarella\altaffilmark{4},
S.~Veilleux\altaffilmark{5},
G.~Bothun\altaffilmark{6},
H.~Inami\altaffilmark{3},
P.~N.~Appleton\altaffilmark{7},
A.~S.~Evans\altaffilmark{8},
S.~Haan\altaffilmark{3},
J.~A.~Marshall\altaffilmark{3},
D.~B.~Sanders\altaffilmark{9},
S.~Stierwalt\altaffilmark{3},
and J.~A.~Surace\altaffilmark{3}
}

\altaffiltext{1}{Department of Physics and Institute of Theoretical and Computational Physics, University of Crete, GR-71003, Heraklion, Greece: tanio@physics.uoc.gr}
\altaffiltext{2}{IESL/Foundation for Research and Technology - Hellas,  GR-71110, Heraklion, Greece and Chercheur Associ\'e, Observatoire de  Paris, F-75014, Paris, France}
\altaffiltext{3}{Spitzer Science Center, Caltech, MS 220-6, Pasadena, CA 91125}
\altaffiltext{4}{Infrared Processing \& Analysis Center, MS 100-22, California Institute of Technology, Pasadena, CA 91125}
\altaffiltext{5}{Department of Physics and Astronomy, University of New York at Stony Brook, NY 11794-3800}
\altaffiltext{6}{Department of Physics, University of Oregon, Eugene, OR 97403}
\altaffiltext{7}{NASA Herschel Science Center, IPAC, MS 100-22, Caltech, Pasadena, CA 91125}
\altaffiltext{8}{Deptartment of Astronomy, 530 McCormick Road, University of Virginia, Charlottesville, VA 22904}
\altaffiltext{9}{Institute for Astronomy, University of Hawaii, 2680 Woodlawn Drive, Honolulu, HI 96822}

\begin{abstract}

We present an analysis of the extended mid-infrared (MIR) emission of the Great Observatories All-Sky LIRG Survey (GOALS) sample based on  $5-15\,\micron$ low resolution spectra obtained with the Infrared Spectrograph on \textit{Spitzer}. We calculate the fraction of extended emission as a function of wavelength for the galaxies in the sample, $FEE_\lambda$, defined as the fraction of the emission which originates outside of the unresolved component of a source at a given distance. We find that the $FEE_\lambda$ varies from one galaxy to another, but we can identify three general types of $FEE_\lambda$: one where $FEE_\lambda$ is constant, one where features due to emission lines and polycyclic aromatic hydrocarbons (PAH) appear more extended than the continuum, and a third which is characteristic of sources with deep silicate absorption at 9.7$\,\micron$. More than 30\% of the galaxies have a median $FEE_\lambda$ larger than 0.5, implying that at least half of their MIR emission is extended. Luminous Infrared Galaxies (LIRGs) display a wide range of $FEE$ in their warm dust continuum ($0\,\lesssim\,FEE_{13.2\mu m}\,\lesssim\,0.85$). The large values of $FEE_{13.2\mu m}$ that we find in many LIRGs suggest that the extended component of their MIR continuum emission originates in scales up to $10\,$kpc, and may contribute as much as the nuclear region to their total MIR luminosity. The mean size of the LIRG cores at 13.2$\,\micron$ is 2.6\,kpc. However, once the IR luminosity of the systems reaches the threshold of $\LIR\,\sim\,10^{11.8}\,\Lsun$, slightly below the regime of Ultra-luminous Infrared Galaxies (ULIRGs), all sources become clearly more compact, with $FEE_{13.2\mu m}\,\lesssim\,0.2$, and their cores are unresolved. Our estimated upper limit for the core size of ULIRGs is less than 1.5\,kpc. Furthermore, our analysis indicates that the compactness of systems with $\LIR\,\gtrsim\,10^{11.25}\,\Lsun$ strongly increases in those classified as mergers in their final stage of interaction. The $FEE_{13.2\mu m}$ is also related to the contribution of an active galactic nucleus (AGN) to the MIR emission. Galaxies which are more AGN-dominated are less extended, independently of their \LIR. We finally find that the extent of the MIR continuum emission is correlated with the far-IR \textit{IRAS} log($f_{60\,\mu m}/f_{100\,\mu m}$) color. This enables us to place a lower limit to the area in a galaxy from where the cold dust emission may originate, a prediction which can be tested soon with the \textit{Herschel} Space Telescope.

\end{abstract}


\section{Introduction}\label{s:intro}

The discovery by the Infrared Astronomical Satellite (IRAS) of luminous and ultra-luminous infrared (IR) galaxies, the so called LIRGs and ULIRGs\footnote{LIRGs are defined as systems displaying an infrared luminosity, \LIR, of: 10$^{11}\,\Lsun\,\leq\,\LIR_{\rm [8-1000\,\mu m]}\,<\,10^{12}\,\Lsun$; ULIRGs: $\LIR_{\rm [8-1000\,\mu m]}\,\geq\,10^{12}\,\Lsun$.}, has opened a new window in extragalactic astrophysics. Over the past 25 years, follow up ground-based and space-born observations of these optically faint systems \citep[see][]{Houck1984} have revealed much about their detailed physical properties as well as their contribution to the integrated energy production in the Universe \citep[see][and references therein]{Sanders1996}. More specifically it has been shown that even though LIRGs and ULIRGs are not very common in the local Universe (\citealt{Soifer1991}), they contribute a substantial fraction of the  energy at z$\,\sim\,1-2$ (\citealt{PG2005}; \citealt{LeFloch2005}; \citealt{Caputi2007}). Furthermore, the more IR luminous systems tend to be more disturbed dynamically, show evidence of merging, and often harbor an active galactic nucleus (AGN). The leap in sensitivity provided by the \textit{Spitzer} Space Telescope (\citealt{Werner2004}) and in particular the availability of deep mid-infrared (MIR) spectroscopy with the Infrared Spectrograph \citep[IRS;][]{Houck2004} enabled the detailed study of the properties of large nearby (i.e., \citealt{Armus2007}, \citealt{Desai2007}, \citealt{Imanishi2008}, \citealt{Farrah2008}; \citealt{PS2010}) and more distant (\citealt{Houck2005}; \citealt[2007]{Yan2005}) samples of (U)LIRGs. It thus became evident that, given the diversity of the MIR spectra of LIRGs and even more of ULIRGs, these systems cannot be grouped in a common class of galaxies either in terms of their MIR and/or far-infrared (FIR) properties. A number of correlations between the MIR colors, strength of features due to polycyclic aromatic hydrocarbons (PAHs), and line emissions, with the dominant source of energy production or interaction stage of galaxies have been explored (\citealt{Armus2007}; \citealt{Desai2007}; \citealt{Veilleux2009}, \citealt{Petric2010}). It appears that systems which are AGN dominated show weak PAH features, which are also correlated with the FIR spectral slope. Systems with greater MIR luminosity display weaker PAH emission, while dust extinction, quantified by the 9.7$\mu$m silicate feature, varies in strength and does not correlate with starburst or AGN dominated systems.

Recently, some studies have suggested that $z\,\sim\,2$ ULIRGs are not just the analogs of local ULIRGs but instead they display MIR spectral features more similar to those seen in local, lower luminosity starburst galaxies and LIRGs (\citealt{Farrah2008}; \citealt{Rigby2008}). Some high-z sub-millimeter galaxies (SMGs) with IR luminosities similar to or greater than ULIRGs also show IR properties different from local ULIRGs (\citealt{Pope2008}; \citealt{Murphy2009}; \citealt{MD2009}). All this evidence, in combination with results obtained from spatially resolved $\Halpha$ imaging and CO and radio maps of SMGs, suggest that the star formation (SF) taking place in ULIRGs and some SMGs at high redshift may be occurring over large areas, extending over several kpc across their disks (\citealt{Bothwell2010}; \citealt{Ivison2010}; \citealt{Alexander2010}). This is in contrast to what is seen in local ULIRGs where the strong bursts of SF are concentrated within the central kpc of galaxies probably due to interactions and mergers (\citealt{Sanders1996}, \citealt{Downes1998}, \citealt{Bryant1999}, \citealt{Soifer2000}), which efficiently drive gas and dust towards their nuclei.

Here we present the first part of an analysis which has as the main goal to quantify the amount of extended MIR emission as a function of wavelength of a large sample of local LIRGs and ULIRGs which can be spatially resolved with \textit{Spitzer}/IRS. This will provide extra evidence on the issue of whether the SF observed in high-redshift ultra-luminous systems has the same MIR characteristics to those observed in disks of lower IR luminosity local systems and it will allow us to examine whether the high-z systems are scaled-up versions in size, SF efficiency and therefore in IR luminosity of local LIRGs or ULIRGs. In this first paper, we study how the extended MIR \textit{continuum} emission relates to global properties, such as the IR luminosity, AGN fraction, merging state, and FIR colors of the galaxies, and derive general trends for the entire sample. In a forthcoming paper (D\'iaz-Santos et al. 2010b, in preparation) we will address how the extended emission varies as function of other 5 to 15$\,\micron$ features, such as PAHs, and emission lines, and how it is linked to the particular physical properties of each galaxy.

The paper is structured as follows: In Section 2 we present our sample, the data reduction, and a summary of the analysis we performed, in order to calculate the extended emission as a function of wavelength for each source. In Section 3 we explore possible correlation of the extended emission with the IR luminosity, stage of interaction, AGN strength and FIR colors for our sample, and discuss some implications for the high-redshift SMG population. The conclusions are presented in Section 4, while in the Appendix we provide more details on the analysis of the spatial profiles and the various tests performed in order to ascertain its robustness.

\section{Observations and Data Analysis}\label{s:irsobs}

\subsection{The sample}\label{ss:sample}

\begin{figure*}
\epsscale{.52}
\plotone{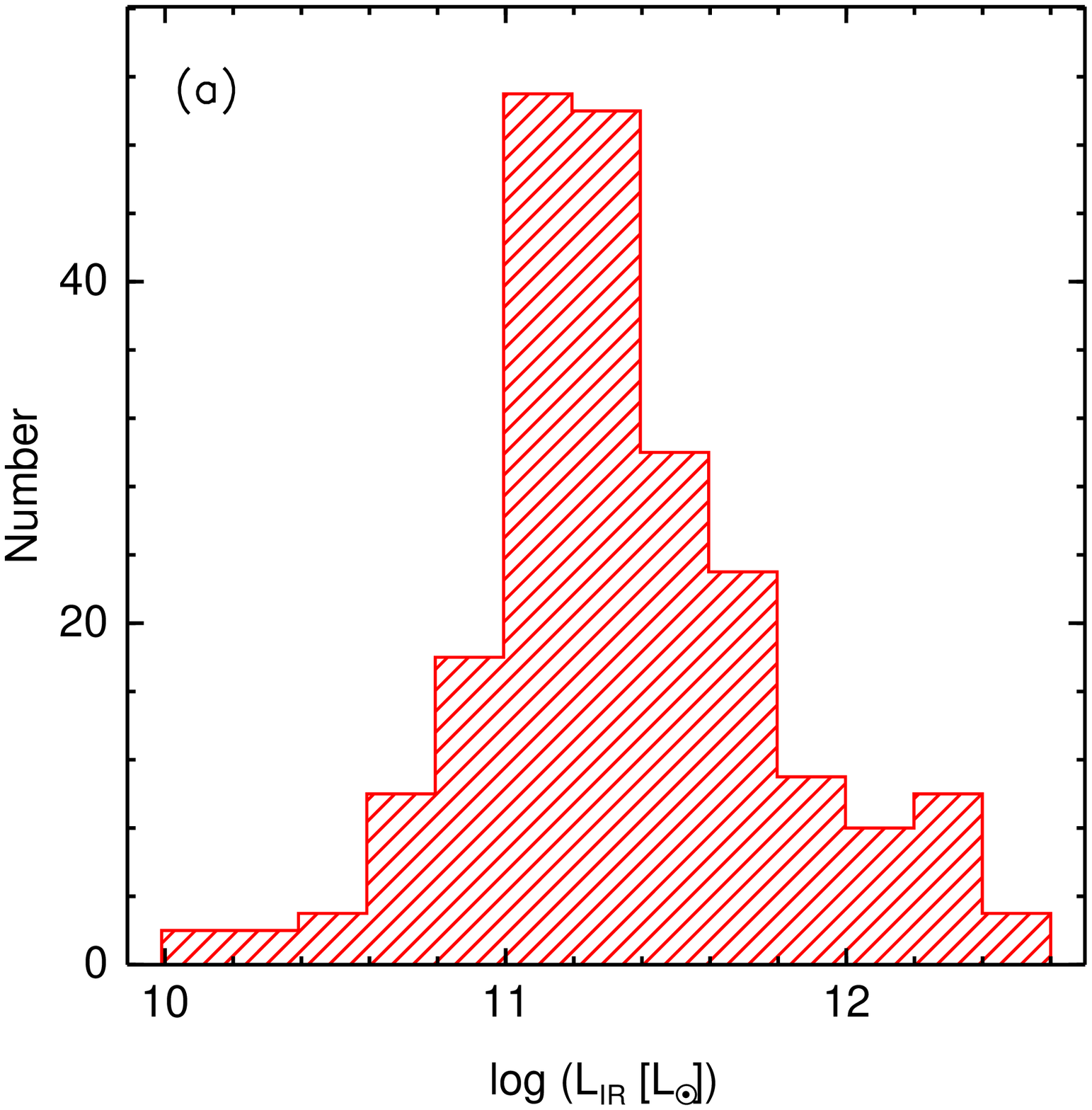}
\epsscale{.58}
\plotone{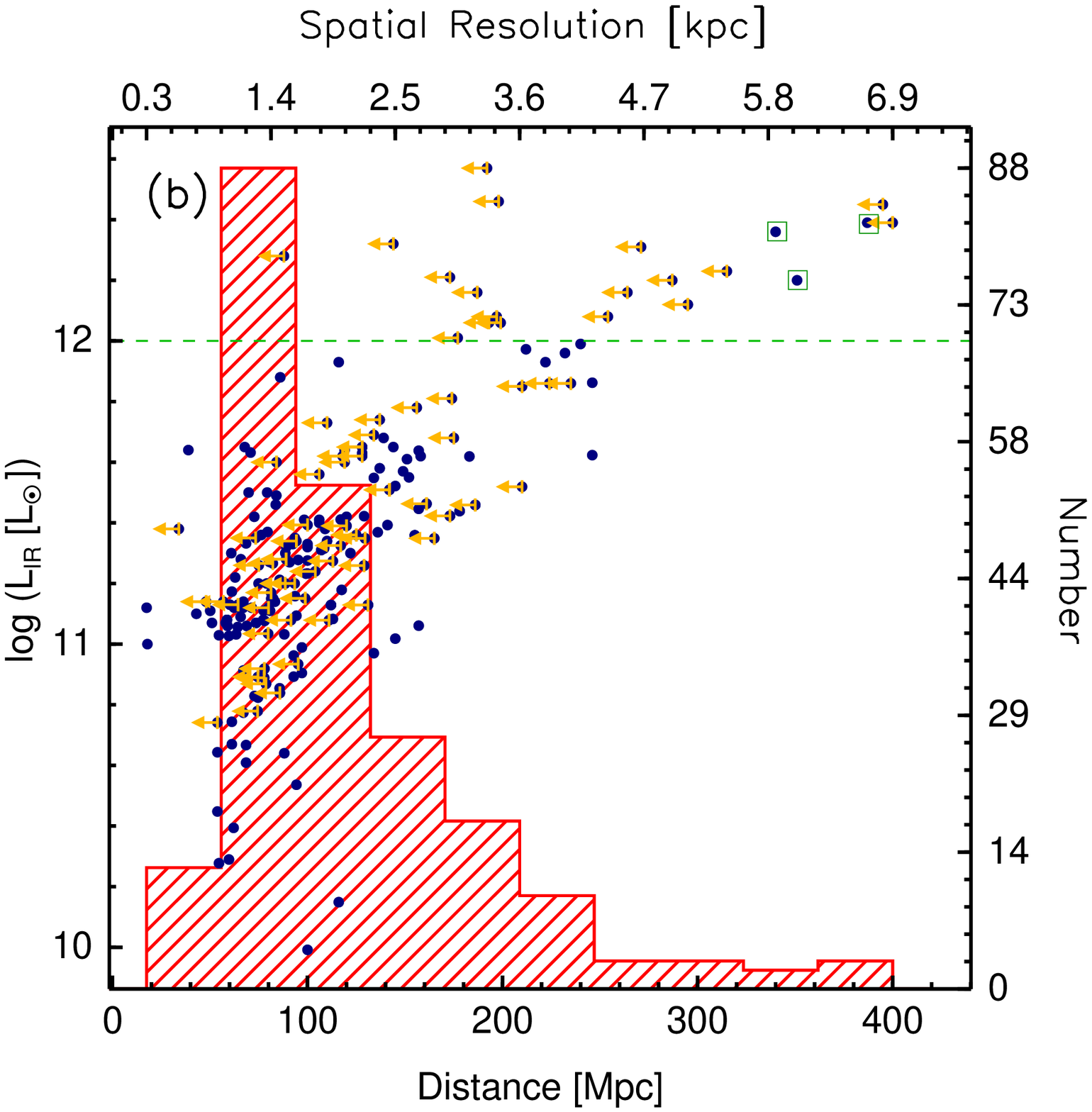}
\vspace{.25cm}
\caption{\footnotesize Histograms of the GOALS galaxy sample. (a) Distribution in IR luminosity; (b) Distribution in distance, indicated on the right y-axis. The upper x-axis displays the projected linear scale that can be resolved at $13.2\,\micron$ at a given distance with IRS. The blue dots are the actual datapoints, with their \LIR\ indicated on the left y-axis of the plot. Upper limits in the x-axis value of a galaxy imply that its ``core'' is unresolved (see Section~\ref{ss:feevslir} and Figure~\ref{f:feevslir}). The green dashed line marks the boundary between LIRGs and ULIRGs. Except of three special cases (green boxes; see also Figure~\ref{f:feevslir}), all ULIRGs have unresolved cores independently of their distance.}\label{f:histos}
\vspace{.5cm}
\end{figure*}

The sample on which we base our analysis is the Great Observatories All-Sky LIRG Survey (GOALS; \citealt{Armus2009}). GOALS comprises a complete, flux-limited sample of galaxies in the local Universe drawn from the Revised Bright Galaxy Sample (RBGS, \citealt{Sanders2003}) selected to be systems in the (U)LIRG luminosity class. \cite{Armus2009} describe in detail the characteristics of the GOALS sample and \cite{Petric2010} presents a comprehensive statistical analysis of the MIR spectral features as probed with \textit{Spitzer}/IRS. Using a number of MIR diagnostics, \cite{Petric2010} estimate the AGN contribution to the MIR luminosity of the systems and, based on their apparent morphology, classify each galaxy into a stage of interaction (from isolated systems to advanced mergers). \textit{Hubble Space Telescope} (\textit{HST}) high spatial resolution optical and near-IR imaging for a fraction of the GOALS sample is presented by \cite{Haan2010}. We refer the reader to these papers for further details. Out of the 291 galaxies (202 systems) included in the GOALS sample \citep[see][]{Armus2009}, a total of 221 are used for this study. This is the number of \textit{individual} galaxies for which IRS staring observations are available. Figure~\ref{f:histos}a presents the distribution of the galaxies as a function of their IR luminosity while Figure~\ref{f:histos}b shows it as a function of the distance. For reference, on the right panel we also indicate the projected linear scales that can be resolved at a given distance at 13.2$\,\micron$ using the IRS $5-15\,\micron$ spectra.

The total IR luminosities of our systems were calculated using their \textit{IRAS} flux densities and luminosity distances \citep[see][for details on the cosmology adopted]{Armus2009}, and following the prescription described in \cite{Sanders1996}. Since the spatial resolution of \textit{Spitzer} is much higher than that of \textit{IRAS}, there are MIPS 24$\,\micron$ \textit{Spitzer} fluxes for 99 individual galaxies within multiple systems. For these,
we distributed the \LIR\ of the system among the galaxies proportionally to their 24$\,\micron$ flux. For those systems without 24$\,\micron$ measurements of individual galaxies, the \LIR\ was evenly distributed among the members of the system. Due to this redistribution of the \LIR, there are now 35 galaxies with IR luminosities lower than 10$^{11}\,\Lsun\,$ in our sample (see Figure~\ref{f:histos}b). In Table~\ref{t:sample} we present the final \LIR\ and distance of the galaxies.

\subsection{Spitzer/IRS Observations}\label{ss:obs}

All galaxies in GOALS have been observed in staring and/or mapping mode with the \textit{Spitzer/IRS} instrument using all instrument modules (SL, LL, SH and LH). They also have IRAC (\citealt{Fazio2004}) and MIPS (\citealt{Rieke2004}) imaging observations at all band-passes \citep{Mazzarella2010}. The analysis presented in this paper is based on IRS SL staring observations covering the 5 to $\sim\,15\,\micron$ wavelength range with a spectral resolution of R$\sim\,60-130$. As mentioned earlier, the aim of the study is to separate and quantify the extended emission of (U)LIRGs from the contribution of the unresolved nuclear component. Thus we limit our study to the shortest wavelengths where the spatial resolution of IRS ($\sim\,3\farcs6$ at $13.2\,\micron$) is sufficient to separate regions of physical scales that range from 0.22\,kpc at the distance of the closest LIRG ($\sim\,12\,$Mpc), to 6.1\,kpc at $\sim\,340\,$Mpc where the farthest ULIRG of the sample is located (see Figure~\ref{f:histos}). The median distance of our galaxy sample is 91\,Mpc, at which the spatial resolution is 1.7\,kpc (also at $13.2\,\micron$). In order to test the validity of some of the results obtained from the IRS SL spectroscopy, we also used the $8\,\micron$ IRAC imaging (see Appendix and \citealt{Mazzarella2010} for details).

\subsection{Data Reduction}\label{ss:datared}

As explained in \cite{Petric2010}, all data were reduced using the S15, S16 and S17 IRS pipelines at the \textit{Spitzer} Science Center\footnote{See http://ssc.spitzer.caltech.edu/irs}\label{foot:pipeline}. The pipelines were changed to modify certain header keywords, produce new flat-fields, and lower SL and LL fringes by 1-20\%, provide better treatment of ramp slopes.
The changes made between the pipelines should not systematically alter the measurements presented here. The reduction includes ramp fitting, dark sky subtraction, droop correction, linearity correction and wavelength and flux calibration.

The backgrounds in the high resolution data were subtracted for all objects with dedicated sky observations ($\sim\,$60\% of the sources). For the low-resolution data without dedicated background observations, off-source nods were used for sky subtraction. Large objects in PID 30323 had dedicated background pointings. Bad pixel mask files were combined such that the final masks flagged all the individual bad pixels, if they were marked as bad in one individual exposure. Basic calibrated data sets (BCDs) for each nod were combined by determining the median if more than 5 BCDs were available, otherwise the average was used with 3-sigma clipping.

Each nod was extracted with SPICE\footnote{See http://ssc.spitzer.caltech.edu/postbcd/doc/spice.pdf} using the standard extraction aperture and point-source calibration. Nod 1 and nod 2 were compared with each other, and pixels were flagged if the difference between nod 1 and nod 2 exceeded 30\%, and when adjacent pixels within the same nod differed by more than 30\% (due to a cosmic ray or hot pixel).


\subsection{Analysis}\label{s:analysis}

In this subsection we briefly summarize how, using $5-15\,\micron$ \textit{Spitzer}/IRS SL staring observations, we can estimate the fraction of the extended emission of the galaxies as a function of wavelength ($FEE_{\lambda}$). More technical details of our analysis, as well as information on the specific treatment of the IRS data, are provided in the Appendix.

We commence the analysis with the 2 dimensional co-added files of the GOALS sample reduced as described above \citep[see][]{Petric2010}. We use the exact algorithms of the SSC pipeline, as well as the grid mapping of the slits on the IRS detector identified by the pseudo-rectangles of the ``wavesamp'' file to extract the spatial profiles of each source along the slit as a function of wavelength for each order and nod position. We follow the same approach to obtain the spatial profiles of a standard star (HR7341; PID: 1432) that we use as the reference point spread function (PSF) of an unresolved source.

For each wavelength resolution element, we fit the spatial profiles of the stellar PSF and the galaxy with Gaussian functions in order to calculate their peaks of emission and FWHMs. The FWHM of the Gaussian used to fit the galaxy profile is what we call the galaxy ``core'' size (see below). We then scale the peak of the spatial profile of the stellar PSF to the maximum of the spatial profile of the galaxy and subtract it. What remains is the spatial profile of the extended emission of the galaxy at a given wavelength. This extended emission ($EE_\lambda$) divided by the integrated emission of the galaxy all along the slit, is what we define as the fraction of extended emission as a function of wavelength, $FEE_\lambda$. We refer the reader to the Appendix for a more detailed explanation on its rather challenging calculation and the corrections that have to be applied to obtain reliable values. So formally:

\begin{equation}\label{e:fee}
FEE_\lambda=\frac{EE_\lambda}{E_\lambda(total)}
\end{equation}

\noindent
where $FEE_\lambda$, $EE_\lambda$ and $E_\lambda(total)$ are the fraction of extended emission (ranging from zero to unity), the extended emission, and the total emission of the galaxy at a given wavelength respectively. Note that since we are scaling the stellar PSF to the maximum of the spatial profile of the galaxy, we implicitly assume that its nuclear emission is entirely produced by the central unresolved component. Therefore the values of $EE_\lambda$ and the $FEE_\lambda$ calculated in this manner are lower limits to the true extended emission. We report as the final $FEE_\lambda$ at a given wavelength the average of the values obtained for each one of the two nod positions of the IRS slits. Comparing the difference between the two nods, we calculate the statistical uncertainty of the $FEE_\lambda$. The final uncertainty includes additional sources of error and depends on the details of the target (see Appendix for more information). For reference, the typical uncertainty on the $FEE$ is between 0.05 and 0.15 (see Table~\ref{t:sample}).

Since the size of the IRS slit used projected on the galaxies does not entirely cover the size of the emitting region, we visually inspected their corresponding IRAC 8$\,\micron$ images in order to examine whether the slit was placed systematically along a privilege direction over the disk of the galaxies. We found that, most likely due to the large size of our sample, the position angle of the slit was randomly distributed along different directions on the galaxies and was not oriented systematically along bars, spiral arms or regions of enhanced star formation. As as consequence, variations in the intrinsic inclinations of the disks in more extended systems do not bias our findings.


\section{Results}

The fraction of the extended emission as a function of wavelength, $FEE_\lambda$, calculated for each one of the sources, contains a wealth of information. In the present section we discuss the types of $FEE_\lambda$ we typically see in our sample, as well as the properties of the extended emission in the MIR continuum. For the latter we use as reference the resolution element at 13.2$\,\micron$ ($FEE_{13.2\mu m}$), as it is an area of the spectrum which is devoid of known emission or absorption features. Our findings though are similar for the $5-15\,\micron$ MIR continuum in general\footnote{Note that this does not hold when the emission at $\sim\,24\,\micron$ is examined since the slit width of the IRS LL module is 10$\farcs$5, nearly three times worse than that of the SL (A. Petric, private communication).}. A detailed comparison of the $FEE_\lambda$ for different MIR spectral features (e.g., PAHs, emission lines or the 9.7$\,\micron$ silicate absorption feature) and the correlation of their ratios with the characteristics of each galaxy (e.g., nuclear and total extinction, presence of an AGN) will be addressed in a companion paper (D\'iaz-Santos et al. 2010b, in preparation). In addition, we construct spectra for the nuclear and extended components of galaxies as a function of the $FEE_\lambda$ types (see next section). Table~\ref{t:sample} presents the $FEE_{13.2\mu m}$ values for the galaxies in our sample, their ``core'' size at the same wavelength (see previous section and Appendix), and their \textit{IRAS} log($f_{60\,\mu m}$/$f_{100\,\mu m}$) ratio, as well as other basic characteristics.

\subsection{Types of $FEE_\lambda$ Profiles}\label{ss:feeclass}

If one ignores projection effects or details on the intrinsic geometry of a source, the spatial profile of a galaxy as a function of wavelength, and consequently the estimated $FEE_\lambda$, depends primarily on the origin of the emission within the galaxy. The MIR continuum emission, PAH features, emission lines, as well as absorption features do not always originate from the same physical regions. Therefore it is reasonable to expect that the $FEE_\lambda$ function varies among the galaxies. We find three different $FEE_\lambda$ types, whose shapes are similar among the galaxies examined here. We exclude from this study 8 sources (4\% of the sample) that appear practically unresolved in the $5-15\,\micron$ range and for which their $FEE_\lambda\,\simeq\,0$. We also exclude 11 galaxies (5\% of the sample) whose $FEE_\lambda$ could not be classified in any of the three types.

\begin{deluxetable*}{lccccccc}
\tabletypesize{\scriptsize}
\tablewidth{0pc}
\tablecaption{\scriptsize Properties Of The Sample}
\tablehead{\colhead{Galaxy} & \colhead{R.A.} & \colhead{Declination} & \colhead{Distance} & \colhead{log $\LIR$} & \colhead{$FEE_{13.2\mu m}$} & \colhead{Core Size} & \colhead{\textit{IRAS}} \\
\colhead{name} & \colhead{(J2000)} & \colhead{(J2000)} & \colhead{[Mpc]} & \colhead{[\Lsun]} & & \colhead{[kpc]} & \colhead{log($f_{60\,\mu m}$/$f_{100\,\mu m}$)} \\
\colhead{(1)} & \colhead{(2)} & \colhead{(3)} & \colhead{(4)} & \colhead{(5)} & \colhead{(6)} & \colhead{(7)} & \colhead{(8)}}
\startdata 
             NGC0023 & 00h 09m 53.4s & +25\deg\ 55m 27s &   65.2 &   11.12 &    0.67$\,\pm$   0.03 &    2.62$\,\pm$   0.02  &  $-$0.239$\,\pm$  0.004 \\
       MCG-02-01-051 & 00h 18m 50.9s & $-$10\deg\ 22m 36s &  117.5 &   11.18 &    0.28$\,\pm$   0.06 &    2.28$\,\pm$   0.02  &  $-$0.111$\,\pm$  0.007 \\
        ESO350-IG038 & 00h 36m 52.5s & $-$33\deg\ 33m 17s &   89.0 &   11.28 &    0.07$\,\pm$   0.08 &    1.57$\,\pm$   0.01* &   0.135$\,\pm$  0.003 \\
             NGC0232 & 00h 42m 45.8s & $-$23\deg\ 33m 41s &   95.2 &   11.28 &    0.37$\,\pm$   0.06 &    2.09$\,\pm$   0.01  &  \dots                \\
             NGC0232 & 00h 42m 52.8s & $-$23\deg\ 32m 27s &   95.2 &   10.93 &    0.15$\,\pm$   0.07 &    1.79$\,\pm$   0.01* &  \dots                \\
       MCG+12-02-001 & 00h 54m 03.9s & +73\deg\ 05m 05s &   69.8 &   11.50 &    0.42$\,\pm$   0.06 &    1.65$\,\pm$   0.01  &  $-$0.123$\,\pm$  0.006 \\
            NGC0317B & 00h 57m 40.4s & +43\deg\ 47m 32s &   77.8 &   10.89 &    0.19$\,\pm$   0.09 &    1.46$\,\pm$   0.01* &  $-$0.172$\,\pm$  0.005 \\
       MCG-03-04-014 & 01h 10m 08.9s & $-$16\deg\ 51m 10s &  144.0 &   11.65 &    0.53$\,\pm$   0.05 &    4.16$\,\pm$   0.05  &  $-$0.154$\,\pm$  0.007 \\
         ESO244-G012 & 01h 18m 08.2s & $-$44\deg\ 28m 00s &   91.5 &   11.08 &    0.15$\,\pm$   0.06 &    1.68$\,\pm$   0.01* &  $-$0.103$\,\pm$  0.004 \\
         CGCG436-030 & 01h 20m 02.6s & +14\deg\ 21m 42s &  134.0 &   11.69 &    0.02$\,\pm$   0.07 &    2.41$\,\pm$   0.02* &   0.044$\,\pm$  0.008 \\
         ESO353-G020 & 01h 34m 51.3s & $-$36\deg\ 08m 14s &   68.8 &   11.06 &    0.47$\,\pm$   0.06 &    1.72$\,\pm$   0.02  &  $-$0.336$\,\pm$  0.004 \\
         ESO297-G011 & 01h 36m 23.4s & $-$37\deg\ 19m 18s &   74.6 &   10.82 &    0.52$\,\pm$   0.05 &    1.97$\,\pm$   0.03  &  \dots                \\
         ESO297-G011 & 01h 36m 24.1s & $-$37\deg\ 20m 25s &   74.6 &   10.89 &    0.07$\,\pm$   0.07 &    1.35$\,\pm$   0.01* &  \dots                \\
     IRASF01364-1042 & 01h 38m 52.8s & $-$10\deg\ 27m 12s &  210.0 &   11.85 &    0.01$\,\pm$   0.08 &    3.79$\,\pm$   0.17* &  $-$0.017$\,\pm$  0.008 \\
     IRASF01417+1651 & 01h 44m 30.6s & +17\deg\ 06m 09s &  119.0 &   11.64 &    0.45$\,\pm$   0.05 &    2.74$\,\pm$   0.22  &  $-$0.033$\,\pm$  0.005 \\
             NGC0695 & 01h 51m 14.4s & +22\deg\ 34m 55s &  139.0 &   11.68 &    0.73$\,\pm$   0.03 &    6.69$\,\pm$   0.09  &  $-$0.252$\,\pm$  0.006 \\
            UGC01385 & 01h 54m 53.8s & +36\deg\ 55m 04s &   79.8 &   11.03 &    0.21$\,\pm$   0.06 &    1.51$\,\pm$   0.02* &  $-$0.123$\,\pm$  0.006 \\
             NGC0838 & 02h 09m 42.8s & $-$10\deg\ 11m 02s &   53.8 &   10.74 &    0.07$\,\pm$   0.06 &    0.96$\,\pm$   0.01* &  \dots                \\
             NGC0838 & 02h 09m 38.7s & $-$10\deg\ 08m 47s &   53.8 &   10.64 &    0.58$\,\pm$   0.04 &    1.65$\,\pm$   0.02  &  \dots                \\
             NGC0838 & 02h 09m 20.9s & $-$10\deg\ 07m 59s &   53.8 &   10.45 &    0.73$\,\pm$   0.03 &    2.70$\,\pm$   0.05  &  \dots                \\
             NGC0828 & 02h 10m 09.5s & +39\deg\ 11m 24s &   76.3 &   11.36 &    0.72$\,\pm$   0.03 &    3.49$\,\pm$   0.02  &  $-$0.344$\,\pm$  0.003 \\
              IC0214 & 02h 14m 05.6s & +05\deg\ 10m 23s &  129.0 &   11.42 &    0.70$\,\pm$   0.04 &    5.82$\,\pm$   0.15  &  $-$0.210$\,\pm$  0.006 \\
             NGC0877 & 02h 17m 59.7s & +14\deg\ 32m 38s &   54.6 &   11.03 &    0.80$\,\pm$   0.06 &    4.79$\,\pm$   1.32  &  \dots                \\
             NGC0877 & 02h 17m 53.3s & +14\deg\ 31m 18s &   54.6 &   10.28 &    0.32$\,\pm$   0.05 &    1.09$\,\pm$   0.02  &  \dots                \\
       MCG+05-06-036 & 02h 23m 22.0s & +32\deg\ 11m 48s &  145.0 &   11.52 &    0.25$\,\pm$   0.06 &    2.93$\,\pm$   0.03  &  \dots                \\
       MCG+05-06-036 & 02h 23m 19.0s & +32\deg\ 11m 18s &  145.0 &   11.02 &    0.53$\,\pm$   0.06 &    3.96$\,\pm$   0.19  &  \dots                \\
            UGC01845 & 02h 24m 08.0s & +47\deg\ 58m 12s &   67.0 &   11.12 &    0.53$\,\pm$   0.05 &    1.88$\,\pm$   0.01  &  $-$0.177$\,\pm$  0.020 \\
             NGC0992 & 02h 37m 25.5s & +21\deg\ 06m 02s &   58.0 &   11.07 &    0.76$\,\pm$   0.03 &    3.26$\,\pm$   0.03  &  $-$0.166$\,\pm$  0.005 \\
            UGC02238 & 02h 46m 17.4s & +13\deg\ 05m 44s &   92.4 &   11.33 &    0.68$\,\pm$   0.03 &    3.64$\,\pm$   0.04  &  $-$0.283$\,\pm$  0.004 \\
     IRASF02437+2122 & 02h 46m 39.1s & +21\deg\ 35m 10s &   98.8 &   11.15 &    0.06$\,\pm$   0.09 &    1.77$\,\pm$   0.04* &  $-$0.053$\,\pm$  0.029 \\
            UGC02369 & 02h 54m 01.8s & +14\deg\ 58m 36s &  136.0 &   11.37 &    0.47$\,\pm$   0.05 &    3.52$\,\pm$   0.03  &  $-$0.142$\,\pm$  0.008 \\
            UGC02608 & 03h 15m 01.5s & +42\deg\ 02m 08s &  100.0 &   11.39 &    0.17$\,\pm$   0.06 &    1.81$\,\pm$   0.02* &  \dots                \\
            UGC02608 & 03h 15m 14.6s & +41\deg\ 58m 50s &  100.0 &    9.99 &    0.49$\,\pm$   0.14 &    3.46$\,\pm$   1.69  &  \dots                \\
             NGC1275 & 03h 19m 48.2s & +41\deg\ 30m 42s &   75.0 &   11.26 &    0.00$\,\pm$   0.06 &    1.27$\,\pm$   0.02* &  $-$0.013$\,\pm$  0.028 \\
     IRASF03217+4022 & 03h 25m 05.4s & +40\deg\ 33m 32s &  100.0 &   11.33 &    0.37$\,\pm$   0.06 &    2.21$\,\pm$   0.05  &  $-$0.163$\,\pm$  0.013 \\
             NGC1365 & 03h 33m 36.4s & $-$36\deg\ 08m 25s &   17.9 &   11.00 &    0.48$\,\pm$   0.04 &    0.43$\,\pm$   0.01  &  $-$0.245$\,\pm$  0.000 \\
     IRASF03359+1523 & 03h 38m 47.1s & +15\deg\ 32m 54s &  152.0 &   11.55 &    0.64$\,\pm$   0.04 &    6.15$\,\pm$   0.68  &  $-$0.086$\,\pm$  0.008 \\
         CGCG465-012 & 03h 54m 07.7s & +15\deg\ 59m 24s &   94.3 &   10.54 &    0.71$\,\pm$   0.04 &    4.52$\,\pm$   0.21  &  \dots                \\
         CGCG465-012 & 03h 54m 16.0s & +15\deg\ 55m 43s &   94.3 &   11.09 &    0.57$\,\pm$   0.04 &    2.61$\,\pm$   0.05  &  \dots                \\
      IRAS03582+6012 & 04h 02m 33.0s & +60\deg\ 20m 41s &  131.0 &   11.13 &    0.05$\,\pm$   0.10 &    2.30$\,\pm$   0.02* &  \dots                \\
      IRAS03582+6012 & 04h 02m 32.0s & +60\deg\ 20m 38s &  131.0 &   11.13 &    0.48$\,\pm$   0.05 &    3.39$\,\pm$   0.05  &  \dots                \\
            UGC02982 & 04h 12m 22.7s & +05\deg\ 32m 49s &   74.9 &   11.20 &    0.70$\,\pm$   0.03 &    3.11$\,\pm$   0.08  &  $-$0.302$\,\pm$  0.008 \\
         ESO420-G013 & 04h 13m 49.7s & $-$32\deg\ 00m 25s &   51.0 &   11.07 &    0.21$\,\pm$   0.07 &    0.98$\,\pm$   0.01  &  $-$0.184$\,\pm$  0.002 \\
             NGC1572 & 04h 22m 42.8s & $-$40\deg\ 36m 03s &   88.6 &   11.30 &    0.41$\,\pm$   0.06 &    2.09$\,\pm$   0.02  &  $-$0.321$\,\pm$  0.003 \\
      IRAS04271+3849 & 04h 30m 33.1s & +38\deg\ 55m 47s &   80.8 &   11.11 &    0.36$\,\pm$   0.05 &    1.74$\,\pm$   0.02  &  $-$0.200$\,\pm$  0.030 \\
             NGC1614 & 04h 33m 60.0s & $-$08\deg\ 34m 46s &   67.8 &   11.65 &    0.30$\,\pm$   0.05 &    1.39$\,\pm$   0.01  &  $-$0.029$\,\pm$  0.006 \\
            UGC03094 & 04h 35m 33.8s & +19\deg\ 10m 18s &  106.0 &   11.41 &    0.71$\,\pm$   0.03 &    4.53$\,\pm$   0.07  &  $-$0.306$\,\pm$  0.010 \\
        ESO203-IG001 & 04h 46m 49.6s & $-$48\deg\ 33m 30s &  235.0 &   11.86 &    0.00$\,\pm$   0.07 &    4.16$\,\pm$   0.03* &   0.037$\,\pm$  0.006 \\
       MCG-05-12-006 & 04h 52m 05.0s & $-$32\deg\ 59m 26s &   81.3 &   11.17 &    0.08$\,\pm$   0.07 &    1.48$\,\pm$   0.02* &  $-$0.082$\,\pm$  0.004 \\
             NGC1797 & 05h 07m 44.8s & $-$08\deg\ 01m 08s &   63.4 &   11.03 &    0.29$\,\pm$   0.06 &    1.26$\,\pm$   0.01  &  $-$0.125$\,\pm$  0.005 \\
         CGCG468-002 & 05h 08m 19.7s & +17\deg\ 21m 47s &   77.9 &   10.92 &    0.04$\,\pm$   0.09 &    1.38$\,\pm$   0.02* &  \dots                \\
         CGCG468-002 & 05h 08m 21.2s & +17\deg\ 22m 08s &   77.9 &   10.92 &    0.22$\,\pm$   0.07 &    1.51$\,\pm$   0.01  &  \dots                \\
      IRAS05083+2441 & 05h 11m 25.9s & +24\deg\ 45m 18s &   99.2 &   11.23 &    0.38$\,\pm$   0.06 &    2.24$\,\pm$   0.01  &  $-$0.082$\,\pm$  0.024 \\
     IRASF05081+7936 & 05h 16m 46.4s & +79\deg\ 40m 13s &  240.0 &   11.99 &    0.43$\,\pm$   0.07 &    6.01$\,\pm$   0.09  &  $-$0.263$\,\pm$  0.006 \\
      IRAS05129+5128 & 05h 16m 56.0s & +51\deg\ 31m 57s &  120.0 &   11.42 &    0.22$\,\pm$   0.06 &    2.38$\,\pm$   0.02  &  $-$0.049$\,\pm$  0.018 \\
     IRASF05189-2524 & 05h 21m 01.4s & $-$25\deg\ 21m 46s &  187.0 &   12.16 &    0.02$\,\pm$   0.09 &    3.24$\,\pm$   0.04* &   0.049$\,\pm$  0.003 \\
     IRASF05187-1017 & 05h 21m 06.5s & $-$10\deg\ 14m 46s &  122.0 &   11.30 &    0.23$\,\pm$   0.09 &    2.38$\,\pm$   0.06  &  $-$0.174$\,\pm$  0.004 \\
      IRAS05223+1908 & 05h 25m 16.7s & +19\deg\ 10m 47s &  128.0 &   11.65 &    0.00$\,\pm$   0.06 &    2.16$\,\pm$   0.01* &  $-$0.299$\,\pm$  0.055 \\
       MCG+08-11-002 & 05h 40m 43.7s & +49\deg\ 41m 41s &   83.7 &   11.46 &    0.30$\,\pm$   0.07 &    1.69$\,\pm$   0.01  &  $-$0.248$\,\pm$  0.003 \\
             NGC1961 & 05h 42m 04.6s & +69\deg\ 22m 42s &   59.0 &   11.06 &    0.76$\,\pm$   0.02 &    3.10$\,\pm$   0.11  &  $-$0.513$\,\pm$  0.005 \\
            UGC03351 & 05h 45m 48.0s & +58\deg\ 42m 03s &   65.8 &   11.28 &    0.74$\,\pm$   0.03 &    3.02$\,\pm$   0.12  &  $-$0.315$\,\pm$  0.002 \\
      IRAS05442+1732 & 05h 47m 11.2s & +17\deg\ 33m 46s &   80.5 &   11.27 &    0.25$\,\pm$   0.06 &    1.56$\,\pm$   0.01  &  $-$0.104$\,\pm$  0.009 \\
     IRASF06076-2139 & 06h 09m 45.7s & $-$21\deg\ 40m 24s &  165.0 &   11.35 &    0.05$\,\pm$   0.11 &    2.96$\,\pm$   0.03* &  $-$0.120$\,\pm$  0.010 \\
            UGC03410 & 06h 14m 29.6s & +80\deg\ 26m 59s &   59.7 &   11.03 &    0.76$\,\pm$   0.03 &    3.26$\,\pm$   0.11  &  \dots                \\
            UGC03410 & 06h 13m 57.9s & +80\deg\ 28m 34s &   59.7 &   10.29 &    0.74$\,\pm$   0.02 &    2.88$\,\pm$   0.09  &  \dots                
\enddata
\end{deluxetable*}

\setcounter{table}{0}
\begin{deluxetable*}{lccccccc}
\tabletypesize{\scriptsize}
\tablewidth{0pc}
\tablecaption{\scriptsize Properties Of The Sample}
\tablehead{\colhead{Galaxy} & \colhead{R.A.} & \colhead{Declination} & \colhead{Distance} & \colhead{log $\LIR$} & \colhead{$FEE_{13.2\mu m}$} & \colhead{Core Size} & \colhead{\textit{IRAS}} \\
\colhead{name} & \colhead{(J2000)} & \colhead{(J2000)} & \colhead{[Mpc]} & \colhead{[\Lsun]} & & \colhead{[kpc]} & \colhead{log($f_{60\,\micron}$/$f_{100\,\micron}$)} \\
\colhead{(1)} & \colhead{(2)} & \colhead{(3)} & \colhead{(4)} & \colhead{(5)} & \colhead{(6)} & \colhead{(7)} & \colhead{(8)}}
\startdata 
             NGC2146 & 06h 18m 37.8s & +78\deg\ 21m 24s &   17.5 &   11.12 &    0.65$\,\pm$   0.03 &    0.62$\,\pm$   0.01  &  $-$0.122$\,\pm$  0.001 \\
        ESO255-IG007 & 06h 27m 21.7s & $-$47\deg\ 10m 36s &  173.0 &   11.42 &    0.17$\,\pm$   0.08 &    3.32$\,\pm$   0.02  &  \dots                \\
        ESO255-IG007 & 06h 27m 22.6s & $-$47\deg\ 10m 47s &  173.0 &   11.42 &    0.52$\,\pm$   0.06 &    4.96$\,\pm$   0.19  &  \dots                \\
        ESO255-IG007 & 06h 27m 23.1s & $-$47\deg\ 11m 02s &  173.0 &   11.42 &    0.27$\,\pm$   0.12 &    3.56$\,\pm$   0.22  &  \dots                \\
         ESO557-G002 & 06h 31m 47.2s & $-$17\deg\ 37m 16s &   93.6 &   11.16 &    0.35$\,\pm$   0.07 &    1.98$\,\pm$   0.02  &  $-$0.151$\,\pm$  0.008 \\
            UGC03608 & 06h 57m 34.4s & +46\deg\ 24m 10s &   94.3 &   11.34 &    0.28$\,\pm$   0.07 &    1.79$\,\pm$   0.01* &  $-$0.148$\,\pm$  0.005 \\
     IRASF06592-6313 & 06h 59m 40.2s & $-$63\deg\ 17m 52s &  104.0 &   11.24 &    0.12$\,\pm$   0.07 &    1.84$\,\pm$   0.02* &  $-$0.117$\,\pm$  0.006 \\
     IRASF07027-6011 & 07h 03m 28.5s & $-$60\deg\ 16m 43s &  141.0 &   11.39 &    0.36$\,\pm$   0.08 &    3.09$\,\pm$   0.08  &  $-$0.141$\,\pm$  0.006 \\
             NGC2342 & 07h 09m 18.1s & +20\deg\ 38m 10s &   78.0 &   11.08 &    0.31$\,\pm$   0.06 &    1.54$\,\pm$   0.02  &  $-$0.494$\,\pm$  0.003 \\
      IRAS07251-0248 & 07h 27m 37.6s & $-$02\deg\ 54m 54s &  400.0 &   12.39 &    0.07$\,\pm$   0.12 &    7.51$\,\pm$   0.08* &   0.009$\,\pm$  0.022 \\
             NGC2388 & 07h 29m 04.6s & +33\deg\ 51m 38s &   62.1 &   10.39 &    0.29$\,\pm$   0.05 &    1.23$\,\pm$   0.01  &  $-$0.167$\,\pm$  0.003 \\
       MCG+02-20-003 & 07h 35m 43.4s & +11\deg\ 42m 34s &   72.8 &   10.83 &    0.29$\,\pm$   0.06 &    1.44$\,\pm$   0.01  &  $-$0.153$\,\pm$  0.004 \\
      IRAS08355-4944 & 08h 37m 01.9s & $-$49\deg\ 54m 29s &  118.0 &   11.62 &    0.20$\,\pm$   0.08 &    2.25$\,\pm$   0.01* &   0.071$\,\pm$  0.042 \\
             NGC2623 & 08h 38m 24.1s & +25\deg\ 45m 16s &   84.1 &   11.60 &    0.11$\,\pm$   0.10 &    1.52$\,\pm$   0.01* &  $-$0.037$\,\pm$  0.002 \\
        ESO432-IG006 & 08h 44m 27.2s & $-$31\deg\ 41m 50s &   74.4 &   10.78 &    0.44$\,\pm$   0.06 &    1.79$\,\pm$   0.02  &  \dots                \\
        ESO432-IG006 & 08h 44m 28.9s & $-$31\deg\ 41m 30s &   74.4 &   10.78 &    0.15$\,\pm$   0.06 &    1.35$\,\pm$   0.02* &  \dots                \\
         ESO60-IG016 & 08h 52m 30.5s & $-$69\deg\ 01m 59s &  210.0 &   11.52 &    0.18$\,\pm$   0.07 &    3.95$\,\pm$   0.03* &  $-$0.030$\,\pm$  0.010 \\
     IRASF08572+3915 & 09h 00m 25.3s & +39\deg\ 03m 53s &  264.0 &   12.16 &    0.00$\,\pm$   0.07 &    4.70$\,\pm$   0.01* &   0.185$\,\pm$  0.014 \\
      IRAS09022-3615 & 09h 04m 12.7s & $-$36\deg\ 27m 01s &  271.0 &   12.31 &    0.11$\,\pm$   0.12 &    5.09$\,\pm$   0.04* &   0.021$\,\pm$  0.014 \\
     IRASF09111-1007 & 09h 13m 36.5s & $-$10\deg\ 19m 29s &  246.0 &   11.86 &    0.20$\,\pm$   0.08 &    4.81$\,\pm$   0.06  &  \dots                \\
     IRASF09111-1007 & 09h 13m 38.9s & $-$10\deg\ 19m 19s &  246.0 &   11.62 &    0.34$\,\pm$   0.08 &    5.19$\,\pm$   0.17  &  \dots                \\
            UGC04881 & 09h 15m 55.5s & +44\deg\ 19m 57s &  178.0 &   11.44 &    0.33$\,\pm$   0.08 &    3.72$\,\pm$   0.12  &  \dots                \\
            UGC04881 & 09h 15m 54.7s & +44\deg\ 19m 50s &  178.0 &   11.44 &    0.47$\,\pm$   0.06 &    4.12$\,\pm$   0.18  &  \dots                \\
            UGC05101 & 09h 35m 51.6s & +61\deg\ 21m 11s &  177.0 &   12.01 &    0.15$\,\pm$   0.07 &    3.30$\,\pm$   0.01* &  $-$0.232$\,\pm$  0.003 \\
       MCG+08-18-013 & 09h 36m 37.2s & +48\deg\ 28m 27s &  117.0 &   11.33 &    0.17$\,\pm$   0.22 &    1.33$\,\pm$   1.69* &  $-$0.171$\,\pm$  0.008 \\
     IRASF09437+0317 & 09h 46m 21.1s & +03\deg\ 04m 16s &   92.9 &   10.89 &    0.65$\,\pm$   0.07 &    3.47$\,\pm$   0.33  &  \dots                \\
     IRASF09437+0317 & 09h 46m 20.3s & +03\deg\ 02m 44s &   92.9 &   10.96 &    0.57$\,\pm$   0.09 &    2.84$\,\pm$   0.31  &  \dots                \\
             NGC3110 & 10h 04m 02.1s & $-$06\deg\ 28m 29s &   79.5 &   11.37 &    0.65$\,\pm$   0.04 &    2.96$\,\pm$   0.04  &  $-$0.295$\,\pm$  0.002 \\
     IRASF10038-3338 & 10h 06m 04.7s & $-$33\deg\ 53m 05s &  156.0 &   11.78 &    0.04$\,\pm$   0.10 &    2.73$\,\pm$   0.02* &   0.036$\,\pm$  0.013 \\
     IRASF10173+0828 & 10h 20m 00.2s & +08\deg\ 13m 32s &  224.0 &   11.86 &    0.00$\,\pm$   0.08 &    3.90$\,\pm$   0.19* &  $-$0.019$\,\pm$  0.008 \\
             NGC3221 & 10h 22m 20.0s & +21\deg\ 34m 10s &   65.7 &   11.09 &    0.83$\,\pm$   0.04 &    5.10$\,\pm$   1.06  &  $-$0.386$\,\pm$  0.003 \\
             NGC3256 & 10h 27m 51.3s & $-$43\deg\ 54m 14s &   38.9 &   11.64 &    0.54$\,\pm$   0.04 &    1.05$\,\pm$   0.01  &  $-$0.047$\,\pm$  0.001 \\
         ESO264-G036 & 10h 43m 07.5s & $-$46\deg\ 12m 44s &  100.0 &   11.32 &    0.68$\,\pm$   0.03 &    3.91$\,\pm$   0.14  &  $-$0.343$\,\pm$  0.018 \\
         ESO264-G057 & 10h 59m 01.7s & $-$43\deg\ 26m 25s &   83.3 &   11.14 &    0.40$\,\pm$   0.06 &    1.79$\,\pm$   0.03  &  $-$0.330$\,\pm$  0.026 \\
     IRASF10565+2448 & 10h 59m 18.1s & +24\deg\ 32m 34s &  197.0 &   12.08 &    0.06$\,\pm$   0.09 &    3.61$\,\pm$   0.02* &  $-$0.094$\,\pm$  0.004 \\
       MCG+07-23-019 & 11h 03m 54.0s & +40\deg\ 51m 00s &  158.0 &   11.62 &    0.61$\,\pm$   0.10 &    6.02$\,\pm$   1.13  &  $-$0.208$\,\pm$  0.005 \\
         CGCG011-076 & 11h 21m 12.2s & $-$02\deg\ 59m 02s &  117.0 &   11.41 &    0.26$\,\pm$   0.08 &    2.36$\,\pm$   0.02  &  $-$0.196$\,\pm$  0.013 \\
              IC2810 & 11h 25m 45.1s & +14\deg\ 40m 36s &  157.0 &   11.45 &    0.47$\,\pm$   0.05 &    3.72$\,\pm$   0.08  &  $-$0.224$\,\pm$  0.006 \\
         ESO319-G022 & 11h 27m 54.2s & $-$41\deg\ 36m 51s &   80.0 &   11.12 &    0.07$\,\pm$   0.09 &    1.39$\,\pm$   0.02* &  $-$0.093$\,\pm$  0.009 \\
        ESO440-IG058 & 12h 06m 51.9s & $-$31\deg\ 56m 59s &  112.0 &   11.13 &    0.52$\,\pm$   0.05 &    3.06$\,\pm$   0.05  &  \dots                \\
        ESO440-IG058 & 12h 06m 51.7s & $-$31\deg\ 56m 46s &  112.0 &   11.13 &    0.53$\,\pm$   0.06 &    2.79$\,\pm$   0.22  &  \dots                \\
     IRASF12112+0305 & 12h 13m 46.0s & +02\deg\ 48m 42s &  340.0 &   12.36 &    0.50$\,\pm$   0.07 &    9.83$\,\pm$   0.22: &  $-$0.063$\,\pm$  0.006 \\
             NGC4194 & 12h 14m 09.7s & +54\deg\ 31m 35s &   43.0 &   11.10 &    0.45$\,\pm$   0.05 &    1.04$\,\pm$   0.01  &  $-$0.035$\,\pm$  0.002 \\
         ESO267-G030 & 12h 14m 12.8s & $-$47\deg\ 13m 42s &   97.1 &   10.90 &    0.57$\,\pm$   0.04 &    2.90$\,\pm$   0.06  &  \dots                \\
         ESO267-G030 & 12h 13m 52.3s & $-$47\deg\ 16m 25s &   97.1 &   10.99 &    0.28$\,\pm$   0.06 &    1.89$\,\pm$   0.03  &  \dots                \\
      IRAS12116-5615 & 12h 14m 22.1s & $-$56\deg\ 32m 32s &  128.0 &   11.65 &    0.17$\,\pm$   0.07 &    2.43$\,\pm$   0.02* &  $-$0.122$\,\pm$  0.017 \\
     IRASF12224-0624 & 12h 25m 03.9s & $-$06\deg\ 40m 51s &  125.0 &   11.36 &    0.07$\,\pm$   0.10 &    2.16$\,\pm$   0.08* &  $-$0.133$\,\pm$  0.008 \\
              Mrk231 & 12h 56m 14.3s & +56\deg\ 52m 24s &  192.0 &   12.57 &    0.00$\,\pm$   0.07 &    3.34$\,\pm$   0.01* &   0.015$\,\pm$  0.002 \\
             NGC4922 & 13h 01m 25.3s & +29\deg\ 18m 49s &  111.0 &   11.08 &    0.04$\,\pm$   0.09 &    1.92$\,\pm$   0.01* &  $-$0.072$\,\pm$  0.006 \\
         CGCG043-099 & 13h 01m 50.3s & +04\deg\ 20m 00s &  175.0 &   11.68 &    0.12$\,\pm$   0.09 &    3.30$\,\pm$   0.05* &  $-$0.186$\,\pm$  0.007 \\
       MCG-02-33-098 & 13h 02m 19.7s & $-$15\deg\ 46m 04s &   78.7 &   10.87 &    0.04$\,\pm$   0.09 &    1.37$\,\pm$   0.01* &  \dots                \\
       MCG-02-33-098 & 13h 02m 20.4s & $-$15\deg\ 45m 59s &   78.7 &   10.87 &    0.39$\,\pm$   0.06 &    1.71$\,\pm$   0.04  &  \dots                \\
         ESO507-G070 & 13h 02m 52.4s & $-$23\deg\ 55m 17s &  106.0 &   11.56 &    0.08$\,\pm$   0.11 &    1.90$\,\pm$   0.02* &  $-$0.081$\,\pm$  0.005 \\
      IRAS13052-5711 & 13h 08m 18.7s & $-$57\deg\ 27m 30s &  106.0 &   11.40 &    0.42$\,\pm$   0.05 &    2.51$\,\pm$   0.04  &  $-$0.252$\,\pm$  0.021 \\
              IC0860 & 13h 15m 03.5s & +24\deg\ 37m 07s &   56.8 &   11.14 &    0.16$\,\pm$   0.06 &    1.02$\,\pm$   0.01* &  $-$0.001$\,\pm$  0.003 \\
      IRAS13120-5453 & 13h 15m 06.4s & $-$55\deg\ 09m 22s &  144.0 &   12.32 &    0.18$\,\pm$   0.07 &    2.68$\,\pm$   0.02* &  $-$0.105$\,\pm$  0.002 \\
              VV250a & 13h 15m 35.0s & +62\deg\ 07m 29s &  142.0 &   11.51 &    0.05$\,\pm$   0.07 &    2.51$\,\pm$   0.01* &  \dots                \\
              VV250a & 13h 15m 30.7s & +62\deg\ 07m 45s &  142.0 &   11.51 &    0.13$\,\pm$   0.11 &    2.70$\,\pm$   0.10* &  \dots                \\
            UGC08387 & 13h 20m 35.4s & +34\deg\ 08m 22s &  110.0 &   11.73 &    0.17$\,\pm$   0.06 &    2.11$\,\pm$   0.01  &  $-$0.156$\,\pm$  0.003 \\
             NGC5104 & 13h 21m 23.1s & +00\deg\ 20m 33s &   90.8 &   11.27 &    0.45$\,\pm$   0.06 &    2.17$\,\pm$   0.04  &  $-$0.295$\,\pm$  0.005 \\
       MCG-03-34-064 & 13h 22m 24.5s & $-$16\deg\ 43m 42s &   82.2 &   11.27 &    0.01$\,\pm$   0.07 &    1.40$\,\pm$   0.01* &   0.000$\,\pm$  0.010 \\
             NGC5135 & 13h 25m 44.0s & $-$29\deg\ 50m 00s &   60.9 &   11.30 &    0.55$\,\pm$   0.05 &    1.77$\,\pm$   0.02  &  $-$0.264$\,\pm$  0.003 \\
         ESO173-G015 & 13h 27m 23.8s & $-$57\deg\ 29m 21s &   34.0 &   11.38 &    0.25$\,\pm$   0.06 &    0.64$\,\pm$   0.01* &  $-$0.089$\,\pm$  0.003 \\
              IC4280 & 13h 32m 53.4s & $-$24\deg\ 12m 25s &   82.4 &   11.15 &    0.76$\,\pm$   0.02 &    4.40$\,\pm$   0.08  &  $-$0.307$\,\pm$  0.017 \\
             NGC5256 & 13h 38m 17.3s & +48\deg\ 16m 32s &  129.0 &   11.26 &    0.36$\,\pm$   0.06 &    2.66$\,\pm$   0.02  &  \dots                
\enddata
\end{deluxetable*}

\setcounter{table}{0}
\begin{deluxetable*}{lccccccc}
\tabletypesize{\scriptsize}
\tablewidth{0pc}
\tablecaption{\scriptsize Properties Of The Sample}
\tablehead{\colhead{Galaxy} & \colhead{R.A.} & \colhead{Declination} & \colhead{Distance} & \colhead{log $\LIR$} & \colhead{$FEE_{13.2\mu m}$} & \colhead{Core Size} & \colhead{\textit{IRAS}} \\
\colhead{name} & \colhead{(J2000)} & \colhead{(J2000)} & \colhead{[Mpc]} & \colhead{[\Lsun]} & & \colhead{[kpc]} & \colhead{log($f_{60\,\micron}$/$f_{100\,\micron}$)} \\
\colhead{(1)} & \colhead{(2)} & \colhead{(3)} & \colhead{(4)} & \colhead{(5)} & \colhead{(6)} & \colhead{(7)} & \colhead{(8)}}
\startdata 
             NGC5256 & 13h 38m 17.8s & +48\deg\ 16m 41s &  129.0 &   11.26 &    0.30$\,\pm$   0.23 &    2.47$\,\pm$   0.11* &  \dots                \\
             NGC5257 & 13h 39m 57.7s & +00\deg\ 49m 53s &  108.5 &   11.32 &    0.78$\,\pm$   0.03 &    9.60$\,\pm$   1.61  &  \dots                \\
             NGC5257 & 13h 39m 52.9s & +00\deg\ 50m 25s &  108.5 &   11.32 &    0.82$\,\pm$   0.04 &    8.98$\,\pm$   1.29  &  \dots                \\
              Mrk273 & 13h 44m 42.1s & +55\deg\ 53m 13s &  173.0 &   12.21 &    0.00$\,\pm$   0.06 &    3.10$\,\pm$   0.02* &  $-$0.000$\,\pm$  0.002 \\
            UGC08739 & 13h 49m 13.9s & +35\deg\ 15m 26s &   81.4 &   11.15 &    0.48$\,\pm$   0.04 &    1.92$\,\pm$   0.03  &  $-$0.438$\,\pm$  0.004 \\
        ESO221-IG010 & 13h 50m 56.9s & $-$49\deg\ 03m 18s &   62.9 &   11.22 &    0.42$\,\pm$   0.06 &    1.40$\,\pm$   0.01  &  $-$0.231$\,\pm$  0.010 \\
             NGC5331 & 13h 52m 16.2s & +02\deg\ 06m 05s &  155.0 &   11.36 &    0.62$\,\pm$   0.03 &    5.25$\,\pm$   0.10  &  \dots                \\
             NGC5331 & 13h 52m 16.4s & +02\deg\ 06m 30s &  155.0 &   11.36 &    0.62$\,\pm$   0.07 &    5.45$\,\pm$   0.32  &  \dots                \\
             NGC5395 & 13h 58m 38.0s & +37\deg\ 25m 28s &   58.7 &   11.08 &    0.68$\,\pm$   0.04 &    2.48$\,\pm$   0.42  &  \dots                \\
             NGC5395 & 13h 58m 33.6s & +37\deg\ 27m 12s &   58.7 &   11.08 &    0.30$\,\pm$   0.05 &    1.18$\,\pm$   0.01  &  \dots                \\
         CGCG247-020 & 14h 19m 43.3s & +49\deg\ 14m 11s &  120.0 &   11.39 &    0.04$\,\pm$   0.09 &    2.15$\,\pm$   0.02* &  $-$0.149$\,\pm$  0.008 \\
             NGC5653 & 14h 30m 10.4s & +31\deg\ 12m 55s &   60.2 &   11.13 &    0.76$\,\pm$   0.02 &    3.37$\,\pm$   0.05  &  $-$0.338$\,\pm$  0.002 \\
     IRASF14348-1447 & 14h 37m 38.3s & $-$15\deg\ 00m 24s &  387.0 &   12.39 &    0.56$\,\pm$   0.05 &   12.37$\,\pm$   0.38: &  $-$0.030$\,\pm$  0.009 \\
     IRASF14378-3651 & 14h 40m 59.0s & $-$37\deg\ 04m 32s &  315.0 &   12.23 &    0.07$\,\pm$   0.12 &    5.78$\,\pm$   0.14* &  $-$0.080$\,\pm$  0.012 \\
             NGC5734 & 14h 45m 09.0s & $-$20\deg\ 52m 13s &   67.1 &   10.91 &    0.77$\,\pm$   0.03 &    3.96$\,\pm$   0.10  &  \dots                \\
             NGC5734 & 14h 45m 11.0s & $-$20\deg\ 54m 48s &   67.1 &   10.77 &    0.65$\,\pm$   0.06 &    2.50$\,\pm$   0.16  &  \dots                \\
              VV340a & 14h 57m 00.7s & +24\deg\ 37m 05s &  157.0 &   11.64 &    0.76$\,\pm$   0.02 &    8.89$\,\pm$   0.18  &  \dots                \\
              VV340a & 14h 57m 00.3s & +24\deg\ 36m 24s &  157.0 &   11.06 &    0.67$\,\pm$   0.07 &    6.16$\,\pm$   0.57  &  \dots                \\
               VV705 & 15h 18m 06.1s & +42\deg\ 44m 44s &  183.0 &   11.62 &    0.60$\,\pm$   0.11 &    6.37$\,\pm$   1.35  &  $-$0.045$\,\pm$  0.004 \\
         ESO099-G004 & 15h 24m 58.0s & $-$63\deg\ 07m 29s &  137.0 &   11.74 &    0.20$\,\pm$   0.10 &    2.57$\,\pm$   0.04* &  $-$0.122$\,\pm$  0.039 \\
     IRASF15250+3608 & 15h 26m 59.4s & +35\deg\ 58m 37s &  254.0 &   12.08 &    0.01$\,\pm$   0.07 &    4.53$\,\pm$   0.04* &   0.078$\,\pm$  0.008 \\
             NGC5936 & 15h 30m 00.8s & +12\deg\ 59m 22s &   67.1 &   11.14 &    0.34$\,\pm$   0.04 &    1.33$\,\pm$   0.01  &  $-$0.306$\,\pm$  0.003 \\
              Arp220 & 15h 34m 57.2s & +23\deg\ 30m 11s &   87.9 &   12.28 &    0.06$\,\pm$   0.12 &    1.53$\,\pm$   0.01* &  $-$0.044$\,\pm$  0.001 \\
             NGC5990 & 15h 46m 16.4s & +02\deg\ 24m 55s &   64.4 &   11.13 &    0.26$\,\pm$   0.06 &    1.22$\,\pm$   0.01* &  $-$0.252$\,\pm$  0.005 \\
             NGC6090 & 16h 11m 40.8s & +52\deg\ 27m 27s &  137.0 &   11.58 &    0.52$\,\pm$   0.04 &    3.90$\,\pm$   0.03  &  $-$0.162$\,\pm$  0.005 \\
     IRASF16164-0746 & 16h 19m 11.8s & $-$07\deg\ 54m 02s &  128.0 &   11.62 &    0.14$\,\pm$   0.07 &    2.42$\,\pm$   0.02* &  $-$0.109$\,\pm$  0.016 \\
         CGCG052-037 & 16h 30m 53.3s & +04\deg\ 04m 23s &  116.0 &   10.15 &    0.36$\,\pm$   0.05 &    2.49$\,\pm$   0.02  &  $-$0.205$\,\pm$  0.005 \\
             NGC6156 & 16h 34m 52.6s & $-$60\deg\ 37m 08s &   48.0 &   11.14 &    0.18$\,\pm$   0.05 &    0.86$\,\pm$   0.01* &  $-$0.270$\,\pm$  0.012 \\
        ESO069-IG006 & 16h 38m 11.9s & $-$68\deg\ 26m 08s &  212.0 &   11.97 &    0.44$\,\pm$   0.07 &    5.34$\,\pm$   0.03  &  $-$0.253$\,\pm$  0.010 \\
     IRASF16399-0937 & 16h 42m 40.1s & $-$09\deg\ 43m 13s &  128.0 &   11.63 &    0.40$\,\pm$   0.07 &    2.76$\,\pm$   0.05  &  $-$0.243$\,\pm$  0.034 \\
         ESO453-G005 & 16h 47m 31.1s & $-$29\deg\ 21m 21s &  100.0 &   11.28 &    0.62$\,\pm$   0.05 &    3.21$\,\pm$   0.19  &  $-$0.105$\,\pm$  0.040 \\
             NGC6240 & 16h 52m 58.9s & +02\deg\ 24m 03s &  116.0 &   11.93 &    0.20$\,\pm$   0.06 &    2.26$\,\pm$   0.01  &  $-$0.062$\,\pm$  0.003 \\
     IRASF16516-0948 & 16h 54m 23.7s & $-$09\deg\ 53m 20s &  107.0 &   11.31 &    0.73$\,\pm$   0.06 &    5.40$\,\pm$   0.66  &  $-$0.340$\,\pm$  0.027 \\
             NGC6286 & 16h 58m 31.6s & +58\deg\ 56m 13s &   85.7 &   11.21 &    0.79$\,\pm$   0.02 &    6.66$\,\pm$   0.11  &  \dots                \\
             NGC6286 & 16h 58m 24.0s & +58\deg\ 57m 21s &   85.7 &   10.85 &    0.56$\,\pm$   0.06 &    2.35$\,\pm$   0.12  &  \dots                \\
     IRASF17132+5313 & 17h 14m 20.4s & +53\deg\ 10m 31s &  232.0 &   11.96 &    0.68$\,\pm$   0.04 &    9.89$\,\pm$   0.21  &  $-$0.114$\,\pm$  0.006 \\
     IRASF17138-1017 & 17h 16m 35.7s & $-$10\deg\ 20m 40s &   84.0 &   11.49 &    0.69$\,\pm$   0.03 &    3.64$\,\pm$   0.06  &  $-$0.098$\,\pm$  0.008 \\
     IRASF17207-0014 & 17h 23m 22.0s & +00\deg\ 17m 01s &  198.0 &   12.46 &    0.17$\,\pm$   0.06 &    3.75$\,\pm$   0.03* &  $-$0.050$\,\pm$  0.007 \\
         ESO138-G027 & 17h 26m 43.3s & $-$59\deg\ 55m 55s &   98.3 &   11.41 &    0.24$\,\pm$   0.09 &    1.90$\,\pm$   0.03  &  $-$0.064$\,\pm$  0.024 \\
            UGC11041 & 17h 54m 51.8s & +34\deg\ 46m 34s &   77.5 &   11.11 &    0.61$\,\pm$   0.03 &    2.50$\,\pm$   0.05  &  $-$0.340$\,\pm$  0.003 \\
         CGCG141-034 & 17h 56m 56.6s & +24\deg\ 01m 01s &   93.4 &   11.20 &    0.19$\,\pm$   0.09 &    1.75$\,\pm$   0.02* &  $-$0.228$\,\pm$  0.010 \\
      IRAS17578-0400 & 18h 00m 31.8s & $-$04\deg\ 00m 53s &   68.5 &   11.33 &    0.47$\,\pm$   0.05 &    1.71$\,\pm$   0.02  &  \dots                \\
      IRAS17578-0400 & 18h 00m 34.1s & $-$04\deg\ 01m 44s &   68.5 &   10.61 &    0.54$\,\pm$   0.06 &    1.91$\,\pm$   0.15  &  \dots                \\
      IRAS17578-0400 & 18h 00m 24.3s & $-$04\deg\ 01m 03s &   68.5 &   10.67 &    0.68$\,\pm$   0.04 &    2.77$\,\pm$   0.17  &  \dots                \\
      IRAS18090+0130 & 18h 11m 38.4s & +01\deg\ 31m 40s &  134.0 &   11.55 &    0.53$\,\pm$   0.05 &    3.60$\,\pm$   0.06  &  \dots                \\
      IRAS18090+0130 & 18h 11m 33.4s & +01\deg\ 31m 42s &  134.0 &   10.97 &    0.28$\,\pm$   0.09 &    2.65$\,\pm$   0.12  &  \dots                \\
         CGCG142-034 & 18h 16m 40.7s & +22\deg\ 06m 46s &   88.1 &   11.03 &    0.37$\,\pm$   0.05 &    1.85$\,\pm$   0.03  &  \dots                \\
         CGCG142-034 & 18h 16m 33.8s & +22\deg\ 06m 38s &   88.1 &   10.64 &    0.43$\,\pm$   0.05 &    2.04$\,\pm$   0.05  &  \dots                \\
     IRASF18293-3413 & 18h 32m 41.1s & $-$34\deg\ 11m 26s &   86.0 &   11.88 &    0.50$\,\pm$   0.05 &    2.26$\,\pm$   0.02  &  $-$0.175$\,\pm$  0.003 \\
           NGC6670AB & 18h 33m 34.2s & +59\deg\ 53m 17s &  129.5 &   11.35 &    0.33$\,\pm$   0.05 &    2.68$\,\pm$   0.07  &  \dots                \\
           NGC6670AB & 18h 33m 37.7s & +59\deg\ 53m 22s &  129.5 &   11.35 &    0.21$\,\pm$   0.08 &    2.44$\,\pm$   0.02* &  \dots                \\
              IC4734 & 18h 38m 25.8s & $-$57\deg\ 29m 25s &   73.4 &   11.35 &    0.23$\,\pm$   0.06 &    1.37$\,\pm$   0.01* &  $-$0.256$\,\pm$  0.003 \\
             NGC6701 & 18h 43m 12.5s & +60\deg\ 39m 11s &   62.4 &   11.12 &    0.45$\,\pm$   0.05 &    1.52$\,\pm$   0.01  &  $-$0.300$\,\pm$  0.002 \\
             NGC6786 & 19h 10m 54.0s & +73\deg\ 24m 36s &  113.0 &   11.08 &    0.53$\,\pm$   0.05 &    2.85$\,\pm$   0.08  &  \dots                \\
             NGC6786 & 19h 11m 04.4s & +73\deg\ 25m 32s &  113.0 &   11.27 &    0.14$\,\pm$   0.11 &    2.06$\,\pm$   0.01* &  \dots                \\
        ESO593-IG008 & 19h 14m 31.1s & $-$21\deg\ 19m 06s &  222.0 &   11.93 &    0.53$\,\pm$   0.06 &    6.28$\,\pm$   0.13  &  $-$0.167$\,\pm$  0.003 \\
     IRASF19297-0406 & 19h 32m 22.3s & $-$04\deg\ 00m 01s &  395.0 &   12.45 &    0.09$\,\pm$   0.09 &    7.51$\,\pm$   0.19* &  $-$0.071$\,\pm$  0.024 \\
      IRAS19542+1110 & 19h 56m 35.8s & +11\deg\ 19m 04s &  295.0 &   12.12 &    0.01$\,\pm$   0.08 &    5.43$\,\pm$   0.09* &  $-$0.003$\,\pm$  0.022 \\
         ESO339-G011 & 19h 57m 37.6s & $-$37\deg\ 56m 08s &   88.6 &   11.20 &    0.18$\,\pm$   0.07 &    1.60$\,\pm$   0.01* &  $-$0.192$\,\pm$  0.015 \\
             NGC6907 & 20h 25m 06.6s & $-$24\deg\ 48m 32s &   50.1 &   11.11 &    0.63$\,\pm$   0.04 &    1.69$\,\pm$   0.02  &  $-$0.321$\,\pm$  0.002 \\
       MCG+04-48-002 & 20h 28m 35.1s & +25\deg\ 44m 00s &   64.2 &   11.06 &    0.45$\,\pm$   0.06 &    1.55$\,\pm$   0.02  &  $-$0.243$\,\pm$  0.006 \\
             NGC6926 & 20h 33m 06.1s & $-$02\deg\ 01m 38s &   89.1 &   11.32 &    0.62$\,\pm$   0.05 &    2.49$\,\pm$   0.20  &  $-$0.307$\,\pm$  0.019 \\
      IRAS20351+2521 & 20h 37m 17.7s & +25\deg\ 31m 37s &  151.0 &   11.61 &    0.43$\,\pm$   0.07 &    3.46$\,\pm$   0.06  &  $-$0.179$\,\pm$  0.010 \\
         CGCG448-020 & 20h 57m 24.1s & +17\deg\ 07m 35s &  161.0 &   11.46 &    0.62$\,\pm$   0.04 &    5.68$\,\pm$   0.13  &  \dots                \\
         CGCG448-020 & 20h 57m 24.4s & +17\deg\ 07m 39s &  161.0 &   11.46 &    0.08$\,\pm$   0.08 &    2.91$\,\pm$   0.01* &  \dots                
\enddata
\end{deluxetable*}

\setcounter{table}{0}
\begin{deluxetable*}{lccccccc}
\tabletypesize{\scriptsize}
\tablewidth{0pc}
\tablecaption{\scriptsize Properties Of The Sample}
\tablehead{\colhead{Galaxy} & \colhead{R.A.} & \colhead{Declination} & \colhead{Distance} & \colhead{log $\LIR$} & \colhead{$FEE_{13.2\mu m}$} & \colhead{Core Size} & \colhead{\textit{IRAS}} \\
\colhead{name} & \colhead{(J2000)} & \colhead{(J2000)} & \colhead{[Mpc]} & \colhead{[\Lsun]} & & \colhead{[kpc]} & \colhead{log($f_{60\,\micron}$/$f_{100\,\micron}$)} \\
\colhead{(1)} & \colhead{(2)} & \colhead{(3)} & \colhead{(4)} & \colhead{(5)} & \colhead{(6)} & \colhead{(7)} & \colhead{(8)}}
\startdata 
        ESO286-IG019 & 20h 58m 26.8s & $-$42\deg\ 39m 00s &  193.0 &   12.06 &    0.03$\,\pm$   0.09 &    3.41$\,\pm$   0.01* &   0.073$\,\pm$  0.003 \\
         ESO286-G035 & 21h 04m 11.1s & $-$43\deg\ 35m 36s &   79.1 &   11.20 &    0.60$\,\pm$   0.04 &    2.56$\,\pm$   0.03  &  $-$0.168$\,\pm$  0.006 \\
      IRAS21101+5810 & 21h 11m 29.3s & +58\deg\ 23m 07s &  174.0 &   11.81 &    0.05$\,\pm$   0.08 &    3.20$\,\pm$   0.02* &  $-$0.145$\,\pm$  0.039 \\
        ESO343-IG013 & 21h 36m 10.5s & $-$38\deg\ 32m 42s &   85.8 &   10.84 &    0.49$\,\pm$   0.06 &    2.14$\,\pm$   0.06  &  \dots                \\
        ESO343-IG013 & 21h 36m 10.9s & $-$38\deg\ 32m 32s &   85.8 &   10.84 &    0.01$\,\pm$   0.07 &    1.51$\,\pm$   0.02* &  \dots                \\
             NGC7130 & 21h 48m 19.5s & $-$34\deg\ 57m 04s &   72.7 &   11.42 &    0.35$\,\pm$   0.08 &    1.46$\,\pm$   0.04  &  $-$0.190$\,\pm$  0.002 \\
         ESO467-G027 & 22h 14m 40.0s & $-$27\deg\ 27m 50s &   77.3 &   11.08 &    0.72$\,\pm$   0.03 &    3.57$\,\pm$   0.11  &  $-$0.350$\,\pm$  0.005 \\
         ESO602-G025 & 22h 31m 25.5s & $-$19\deg\ 02m 03s &  110.0 &   11.34 &    0.36$\,\pm$   0.07 &    2.39$\,\pm$   0.02  &  $-$0.250$\,\pm$  0.005 \\
            UGC12150 & 22h 41m 12.2s & +34\deg\ 14m 56s &   93.5 &   11.35 &    0.36$\,\pm$   0.05 &    1.98$\,\pm$   0.03  &  $-$0.289$\,\pm$  0.005 \\
     IRASF22491-1808 & 22h 51m 49.3s & $-$17\deg\ 52m 24s &  351.0 &   12.20 &    0.21$\,\pm$   0.08 &    7.08$\,\pm$   0.28: &   0.077$\,\pm$  0.009 \\
             NGC7469 & 23h 03m 15.6s & +08\deg\ 52m 25s &   70.8 &   11.63 &    0.20$\,\pm$   0.06 &    1.37$\,\pm$   0.01  &  $-$0.109$\,\pm$  0.007 \\
         CGCG453-062 & 23h 04m 56.5s & +19\deg\ 33m 07s &  109.0 &   11.38 &    0.69$\,\pm$   0.03 &    4.52$\,\pm$   0.24  &  $-$0.213$\,\pm$  0.007 \\
        ESO148-IG002 & 23h 15m 46.8s & $-$59\deg\ 03m 15s &  199.0 &   12.06 &    0.02$\,\pm$   0.08 &    3.65$\,\pm$   0.02* &   0.010$\,\pm$  0.004 \\
              IC5298 & 23h 16m 00.7s & +25\deg\ 33m 24s &  119.0 &   11.60 &    0.02$\,\pm$   0.07 &    2.10$\,\pm$   0.01* &  $-$0.122$\,\pm$  0.004 \\
             NGC7591 & 23h 18m 16.2s & +06\deg\ 35m 09s &   71.4 &   11.12 &    0.34$\,\pm$   0.05 &    1.41$\,\pm$   0.02  &  $-$0.276$\,\pm$  0.010 \\
        ESO077-IG014 & 23h 21m 05.4s & $-$69\deg\ 12m 47s &  186.0 &   11.46 &    0.14$\,\pm$   0.11 &    3.46$\,\pm$   0.07* &  \dots                \\
        ESO077-IG014 & 23h 21m 03.7s & $-$69\deg\ 13m 00s &  186.0 &   11.46 &    0.21$\,\pm$   0.09 &    3.60$\,\pm$   0.11  &  \dots                \\
     IRASF23365+3604 & 23h 39m 01.3s & +36\deg\ 21m 08s &  287.0 &   12.20 &    0.09$\,\pm$   0.12 &    5.38$\,\pm$   0.14* &  $-$0.083$\,\pm$  0.011 \\
       MCG-01-60-022 & 23h 42m 00.9s & $-$03\deg\ 36m 54s &  100.0 &   11.23 &    0.43$\,\pm$   0.07 &    2.27$\,\pm$   0.05  &  $-$0.185$\,\pm$  0.007 \\
      IRAS23436+5257 & 23h 46m 05.4s & +53\deg\ 14m 01s &  149.0 &   11.57 &    0.36$\,\pm$   0.09 &    3.13$\,\pm$   0.11  &  $-$0.202$\,\pm$  0.009 \\
             NGC7753 & 23h 47m 04.8s & +29\deg\ 29m 00s &   73.6 &   11.07 &    0.31$\,\pm$   0.12 &    1.42$\,\pm$   0.10  &  \dots                \\
             NGC7753 & 23h 46m 58.6s & +29\deg\ 27m 32s &   73.6 &   11.07 &    0.64$\,\pm$   0.04 &    2.68$\,\pm$   0.07  &  \dots                \\
             NGC7771 & 23h 51m 03.9s & +20\deg\ 09m 01s &   61.2 &   10.74 &    0.56$\,\pm$   0.04 &    1.45$\,\pm$   0.04  &  \dots                \\
             NGC7771 & 23h 51m 24.8s & +20\deg\ 06m 42s &   61.2 &   11.17 &    0.51$\,\pm$   0.04 &    1.66$\,\pm$   0.02  &  \dots                \\
             NGC7771 & 23h 51m 22.6s & +20\deg\ 05m 49s &   61.2 &   10.67 &    0.43$\,\pm$   0.07 &    1.43$\,\pm$   0.07  &  \dots                \\
             Mrk0331 & 23h 51m 26.8s & +20\deg\ 35m 10s &   79.3 &   11.50 &    0.37$\,\pm$   0.07 &    1.77$\,\pm$   0.02  &  $-$0.101$\,\pm$  0.004
\enddata
\tablecomments{\footnotesize (1) Galaxy name. Multiple systems are indicated with the same name but providing the right ascension and declination of the individual galaxies; (2) Right Ascension (J2000); (3) Declination (J2000); (4) Luminosity distance \citep[see][for details]{Armus2009}. The \LIR\ is accurate to within 0.01\,dex (U et al. 2011, in preparation); (5) IR luminosity of the galaxy (see Section~\ref{ss:sample} for details on its calculation); (6) Fraction of extended emission at 13.2$\,\micron$ ($FEE_{13.2\mu m}$); (7) Size (FWHM) of the galaxy core at 13.2$\,\micron$ (see Appendix). The asterisks denote galaxies with unresolved cores (FWHM within 10\% of that of the unresolved stellar PSF). The colons indicate galaxies whose IRS spectra has been contaminated by a close companion galaxy. In this cases, the $FEE_{13.2\mu m}$ and the core size are not reliable; (8) \textit{IRAS} log($f_{60\,\mu m}/f_{100\,\mu m}$) color. Values are given only for sources with a unique \textit{IRAS} color, i.e., not associated to two individual galaxies.}\label{t:sample}
\end{deluxetable*}

\begin{figure*}
\epsscale{1.15}
\plotone{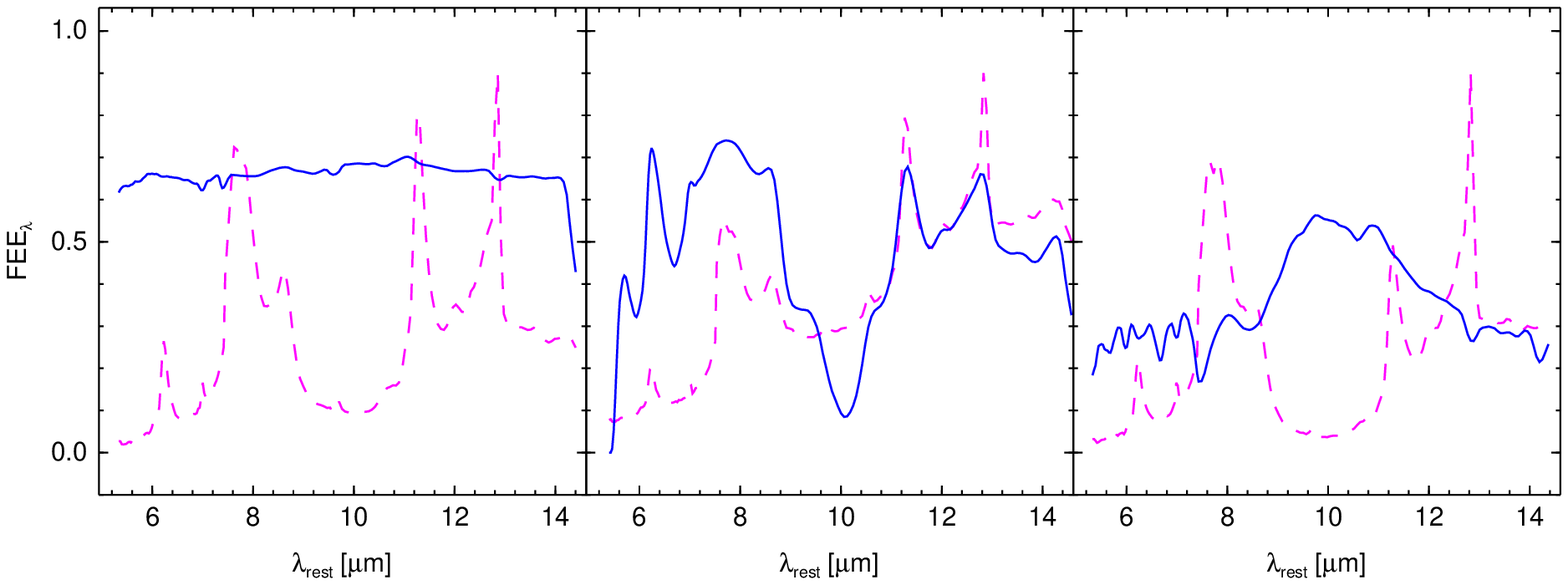}
\vspace{.25cm}
\caption{\footnotesize $FEE_\lambda$ function of 3 galaxies that serve as examples of the 3 types identified in the sample (blue solid line). The spectrum of each galaxy, scaled to arbitrary units, is also plotted for reference (pink dashed line). The galaxies are: NGC~3110 (left), NGC~1365 (center), and MCG+08-11-002 (right). The $FEE_\lambda$ functions have been smoothed with a 4-pixel box to reduce the noise. Left panel: Constant/featureless $FEE_\lambda$. Middle panel: PAH and line extended emission. Right panel: Silicate-extended emission.
}\label{f:feeclass}
\vspace{.5cm}
\end{figure*}

\begin{itemize}

\item Constant/featureless: 111 (50\%) (U)LIRGs display a constant $FEE_\lambda$ across the whole IRS $5-15\,\micron$ wavelength range (see Figure~\ref{f:feeclass}, left panel). No MIR feature appears more extended than another or even than the continuum emission. This implies that there is no differentiation in the spatial distribution of the type of emitting region along the IRS slit. These constant $FEE_\lambda$ functions range between $\sim\,0.1$ up to $\sim\,0.85$ among the galaxies of this type. A representative member of this type is NGC~3110 that, in particular, have a constant $FEE_\lambda$ of $\sim\,0.65$.

\item PAH- and line-extended: 37 galaxies (17\%) of the sample show MIR features which are clearly more extended than the continuum emission (see Figure~\ref{f:feeclass}, central panel). More than $\sim\,20$\% of the flux detected in several PAH features and emission lines is extended, and in some sources it can be as high as 70\%. In these type of galaxies, the bulk of the continuum emission originates from a circumnuclear region more compact than that giving rise to the MIR spectral features. The fact that these two components of emission are spatially decoupled suggests that, contrary to the previous type of $FEE_\lambda$, different physical processes are responsible for the energy production in the nucleus and in the disk. This could be due to the presence of either an AGN or intense starburst in the nucleus producing a strong MIR continuum, compared to a more quiescent, star formation on the disk where older stellar populations may heat the gas and PAH molecules. A representative member of this type is NGC~1365.

\item Silicate-extended: Another 54 (24\%) galaxies of the sample show a rather interesting $FEE_\lambda$ shape in which the maximum peaks around 10$\,\micron$ (see Figure~\ref{f:feeclass}, right panel). Inspecting the $FEE_\lambda$ more closely, it appears that there is an increase between $\sim\,$8 and 12$\,\micron$, around the wavelength range where the 9.7$\,\micron$ silicate absorption feature is present. We attribute this maximum to an extinction effect in these systems. Galaxies containing very obscured nuclei will show little or not at all ``unresolved emission'', at wavelengths dominated by the silicate absorption feature. As mentioned earlier, we always normalize the spatial profile of an unresolved point source to the source profile at every wavelength and subtract it in order to determine the value of $FEE_\lambda$. Consequently, if the nucleus of a galaxy is very extinguished, any residual emission within the 9.7$\mu$m band, when compared to the emission at any other wavelength that is not affected by the absorption feature, would be interpreted as an excess in the fraction of extended emission originating from the outer disk of these galaxies, where the extinction is less extreme. This is reflected in the form of $FEE_\lambda$ we compute, and implies that the silicate strength (i.e., the apparent optical depth) calculated from the integrated spectrum of these galaxies may not be representative of their nuclei \citep[see][]{Spoon2004}. A representative member of this type is MCG+08-11-002.

\end{itemize}

The presence of these types of $FEE_\lambda$, suggest that the mechanism by which local (U)LIRGs produce the MIR emission in their disks and nuclear regions varies from one source to the next. There is a great diversity not only in the integrated spectrum of a LIRG/ULIRG, but also in the spatial distribution of the regions responsible for the formation of the various MIR features, such as PAHs, emission lines, and dust continuum. The $FEE_\lambda$ measured indicates that in some galaxies the emission from PAHs, atomic or molecular lines is more extended than the dust continuum emission, while in others their extent is similar. These variations are clearly related to the location and intensity of the physical process producing the emission detected, such as young star clusters or an AGN. Our results suggest that not all local (U)LIRGs are to be a priori considered as compact, uniform MIR emitters. There has been an interesting such case reported already, VV114, a system with $\LIR\,=\,4\,\times10^{11}\,\Lsun$ harboring an AGN, for which high resolution ground-based MIR imaging \citep{Soifer2001}, as well as $5-15\,\micron$ ISO/CAM spectral maps reveal that nearly 60\% of the MIR emission is extended \citep{LeFloch2002}. This is understood since in these systems the MIR emission is not necessarily dominated by an AGN or a single burst of star formation. Moreover, in the cases of purely starburst-driven (U)LIRGs, the PAH and continuum emission may not be associated to
the same star formation event but instead are probably excited, as a whole, by different stellar populations. For example the \PAHa\ emission is more representative of \textit{recent} star formation event with ages greater than $8-10\,$Myr, while \textit{current} star formation, less than $8-10\,$Myr in age, is better traced by the \NeII\ or MIR continuum emission \citep{DS2010}. In these galaxies, older star formation traced by PAH emission is more extended and is more typical of galactic disks, while recent star formation, associated to MIR continuum emission, is more compact and concentrated towards the nucleus.


\subsection{The \LIR\ Dependence Of The Extended Emission}\label{ss:feevslir}

\begin{figure}
\epsscale{1.15}
\plotone{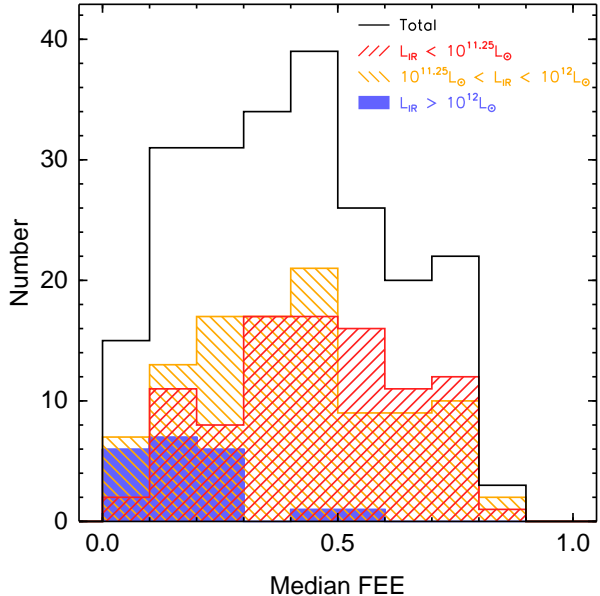}
\vspace{.25cm}
\caption{\footnotesize Histogram of the median $FEE_\lambda$ (calculated over the $5-15\,\micron$ range) for the GOALS sample (black). The red and orange stripped, and the blue solid histograms are the distributions of the median $FEE_\lambda$  for galaxies with $\LIR\,<\,10^{11.25}\,\Lsun$, $10^{11.25}\,\Lsun\,\leq\,\LIR\,<\,10^{12}\,\Lsun$, and $\LIR\,>\,10^{12}\,\Lsun$, respectively.}\label{f:medianfee}
\vspace{.5cm}
\end{figure}

The histogram in Figure~\ref{f:medianfee} presents the distribution of the median $FEE_\lambda$, calculated over the whole $5-15\,\micron$ range, for all the galaxies in our sample (black histogram). 32\% of the galaxies have a median $FEE_\lambda$ larger than 0.5, that is, at least 50\% of the MIR emission of these galaxies is extended. In addition, more than 90\% of the galaxies have a median $FEE_\lambda$ larger than 0.1.

High resolution MIR images of a handful of local (U)LIRGs have revealed rather compact emission originating from a few hundred parsecs around their nuclei (\citealt{Soifer2000}, 2001; \citealt{DS2008}). This trend is also present in our data, even though the physical scales we probe with \textit{Spitzer} are larger. In Figure~\ref{f:medianfee} we also show three histograms of the median $FEE_\lambda$ for galaxies grouped in three ranges of IR luminosity. We note that galaxies with the $\LIR\,<\,10^{11.25}\,\Lsun$ (red histogram) and $10^{11.25}\,\Lsun\,\leq\,\LIR\,<\,10^{12}\,\Lsun$ (orange) display similar median values of their $FEE_\lambda$ distributions. The median of each histogram is 0.46 and 0.39 respectively, which is in agreement with previous studies based on smaller samples of LIRGs \citep{PS2010}. The median values of the low IR luminosity bins are very similar to that found for the median $FEE_\lambda$ distribution of the whole sample of galaxies, 0.40. However, we find that the median of the corresponding distribution for ULIRGs (blue histogram) is only 0.14. A Kolmogorov-Smirnov (K-S) test comparing the two LIRG samples with that of the ULIRG indicates significance levels lower than $1\times10^{-5}$, implying that it is very unlikely that ULIRGs are drawn from the same parent distribution as LIRGs.
Our data therefore suggest that, as a whole, the median $FEE_\lambda$ of ULIRGs is $2-3$ times lower than that of LIRGs.

\begin{figure*}
\epsscale{0.55}
\plotone{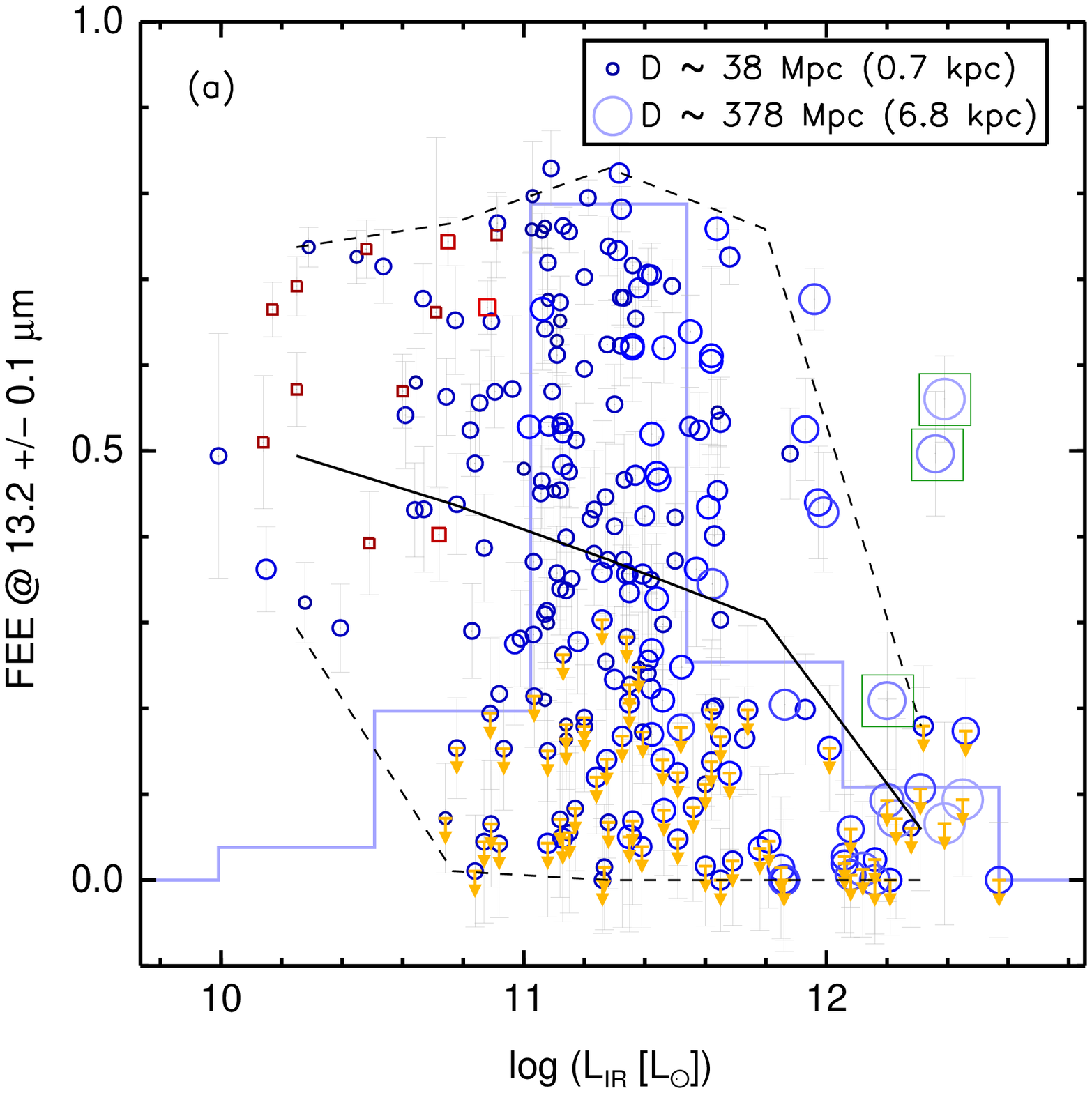}\plotone{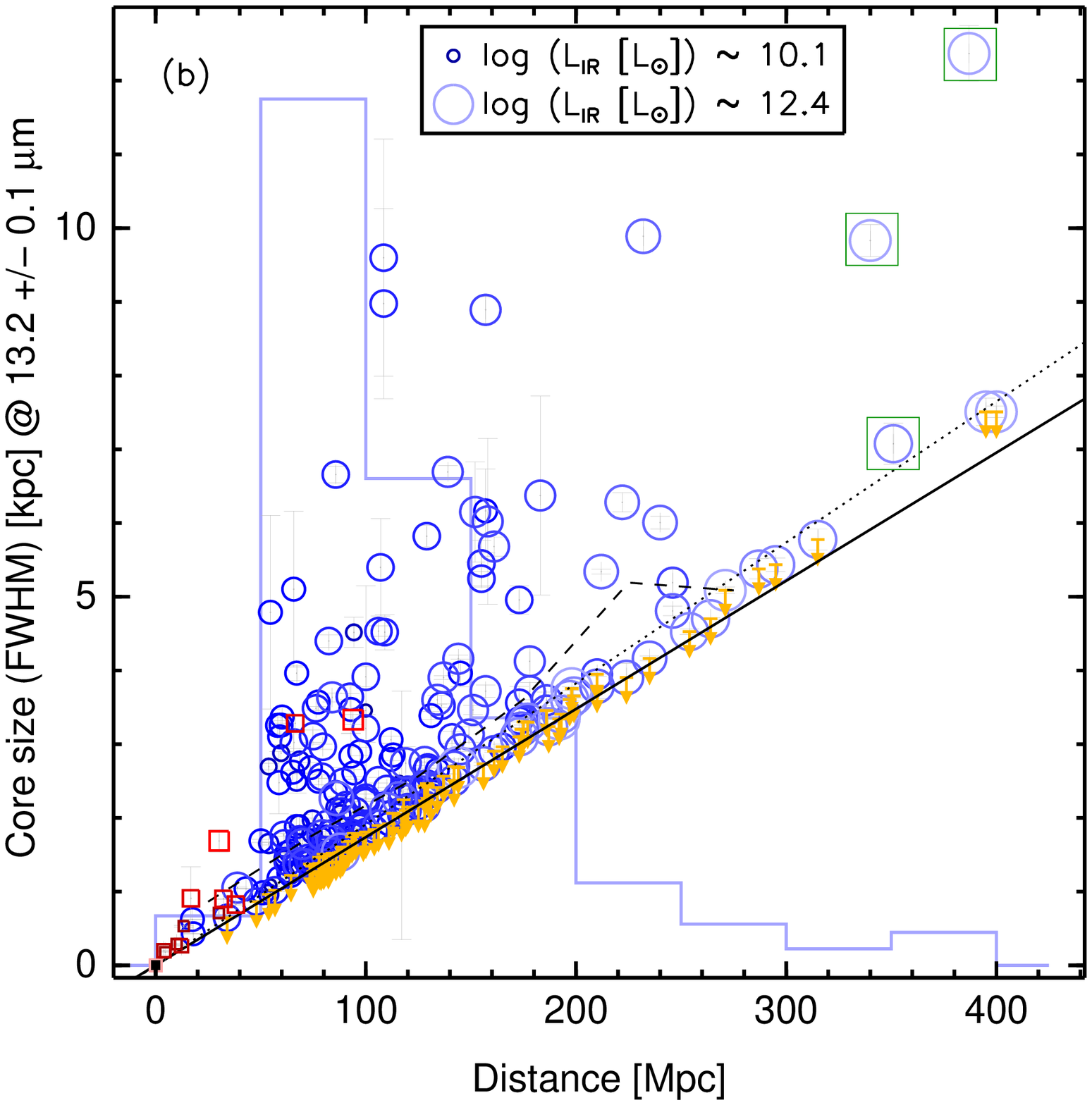}
\vspace{.25cm}
\caption{\footnotesize (a) Plot of the $FEE$ of the continuum at $13.2\,\micron$ as a function of the IR luminosity for the galaxies of our sample (blue circles). The size of the circles scales with the distance to the galaxy. The red squares in the left and right panels are the results obtained for a sub-sample of galaxies taken from the starburst sample of \citet[see text for details]{Brandl2006}. The size of the squares also scales with distance. For reference, the projected linear size of the unresolved component at the given distance is shown in parenthesis. The background faint (blue) line is a normalized histogram of the galaxies at different IR luminosity bins. The solid (black) line is the median of the $FEE_{13.2\mu m}$ at the different luminosity bins. The lower and upper dashed lines are the minimum and maximum $FEE_{13.2\mu m}$ at each bin respectively. The upper limits marked with orange arrows indicate galaxies whose core sizes are unresolved (see next). (b) Plot of the linear size of the galaxy core, measured as the FWHM of the Gaussian function fitted to the spatial profile at $13.2\,\micron$, as a function of distance (see text and Appendix for details). As in (a), the size of the circles and squares scales with the IR luminosity of the source. The faint (blue) line is a normalized histogram of the galaxies at different distance bins. The solid line represents the size of an unresolved source, the FWHM of the stellar PSF, at $13.2\,\micron$ as a function of distance. The black dashed line is the median size of galaxies in each distance bin. The black dotted line indicates the expected location of a galaxy with a core size 10\% larger than that of the unresolved stellar PSF (solid line). The three systems which deviate from the main trend due to contamination of the MIR profile by a companion galaxy are marked with green boxes (see text).}\label{f:feevslir}
\vspace{.5cm}
\end{figure*}

We now explore whether the spatial extent of the MIR continuum of (U)LIRGs is a smooth function of their IR luminosity, or whether there is a certain \LIR\ above which their properties change drastically making their emission substantially more concentrated. The blue circles in Figure~\ref{f:feevslir}a show the $FEE$ of the $13.2\,\micron$ continuum emission ($FEE_{13.2\mu m}$) as a function of the \LIR\ for the galaxies in our sample. The size of the circles scales with their distance. As we can see, LIRGs display a wide range of $FEE_{13.2\mu m}$. They vary from compact systems ($FEE_{13.2\mu m}\,\simeq\,0$) to others which are very extended ($FEE_{13.2\mu m}\,\simeq\,0.85$). No clear trend with \LIR\ is evident over the decade in IR luminosity covered by the LIRGs. However, if we examine the median value of the $FEE_{13.2\mu m}$ calculated in different luminosity bins (black line), this appears to decrease as the \LIR\ of the galaxies increases, going from $\simeq\,0.5$ at $\LIR\,\simeq\,10^{10.3}\,\Lsun\,$ to $\simeq\,0.1$ at $\LIR\,\simeq\,10^{12.3}\,\Lsun\,$. Moreover, for IR luminosities above $\sim\,10^{11.8}\,\Lsun\,$, just below the nominal transition between LIRGs and ULIRGs, the maximum of the $FEE_{13.2\mu m}$ (upper dashed line) drops abruptly, and only a few galaxies have $FEE_{13.2\mu m}$ values larger than 0.2. Indeed, all ULIRGs are unresolved\footnote{There are three systems which deviate from this trend: ESO557-G002, IRAS03359+1523 and IRAS22491-1808. They are interacting systems and their IRAC 8$\,\micron$ images show the presence of nearby companion galaxies affecting the spatial profiles estimated with the IRS slits.
These three systems are marked with green boxes in Figures~\ref{f:histos}b and \ref{f:feevslir}, and they are no longer considered in our analysis.} (see Figure~\ref{f:feevslir}b) and have $FEE_{13.2\mu m}\,\lesssim\,0.2$.


One may consider that the decrease of the median $FEE_{13.2\mu m}$ with \LIR\ is due to the loss of spatial resolution (see Figure~\ref{f:feevslir}b and Figure~\ref{f:histos}b). The more luminous systems tend to be at larger distances and hence their unresolved nuclear component progressively accounts for a larger fraction of their total flux. This bias would affect our ability to measure large values of $FEE_{13.2\mu m}$ for distant sources as they become progressively unresolved. We define as unresolved those galaxies whose ``core'' angular sizes are within 10\% of the stellar PSF, and indicate it as a dotted line in Figure~\ref{f:feevslir}b.
At this point, it is important to differentiate between the core size of a galaxy, used to establish whether it is resolved or not, and its measured $FEE$. While the former is merely a Gaussian fit to the nuclear emission of the galaxy from which a FWHM is obtained, the latter also accounts for possible low surface brightness emission which is probably more extended than the core of the galaxy. As a consequence, sources classified as unresolved in Figure~\ref{f:feevslir}b, may still display non-zero $FEE_{13.2\mu m}$ values in Figure~\ref{f:feevslir}a.

If our ability to measure the $FEE$ and the size of the MIR emitting region of the galaxies were totally dominated by the instrumental resolution, the fraction of unresolved sources would be a monotonically increasing linear function of distance. To examine this, we divided our sample in 3 distance bins at [25-75], [75-125] and [150-250]\,Mpc, each doubling the distance of the previous one, and calculated the fraction of unresolved sources in each one. We find that the corresponding fraction of unresolved sources is 16\% (9/55), 33\% (29/88) and 45\% (18/40). That is, as the distance doubles, the fraction of unresolved sources nearly doubles. However, there are sources up to $\simeq\,240\,$Mpc, for which we can measure a $FEE_{13.2\mum}\,\simeq\,0.7$ independently of their \LIR, implying that we are able to resolve the extended emission at those distances if present.

Hence, the steep drop of the maximum value of $FEE_{13.2\mu m}$ above $\LIR\,\sim\,10^{11.8}\,\Lsun\,$ can not be explained by a progressive increasing of the distance to the galaxies. We note that all 10 ULIRGs found at distances up to 200\,Mpc, where we are still able to  measure large values of $FEE_{13.2\mu m}$, are unresolved and display $FEE_{13.2\mu m}\,<\,0.2$ (see Figure~\ref{f:feevslir}b). However, 68\% (54/80) of LIRGs at similar distances, between 100 and 200\,Mpc, have large $FEEs_{13.2\mu m}$ reaching values up to 0.7. If ULIRGs were drawn from the same parent population as LIRGs, one would expect to resolve around 6 out of these 10 ULIRGs, which is not the case. Therefore, it appears that there is a real threshold in the distribution of the compactness of IR bright galaxies happening at $\LIR\,\sim\,10^{11.8}\,\Lsun$, with higher luminosity systems being more compact.

When we examine only the LIRGs of our sample, we find that, independently of the distance, there are systems with more than 50\% of their MIR continuum emission originating outside the unresolved central region. This implies that the extended emission in these systems is larger than the contribution of the nuclear region to their total MIR luminosity. In fact, we are able to resolve sizes (FWHMs) of LIRG cores up to $\sim\,10\,$kpc, much larger than the spatial resolution at any given distance. However, since our sample is flux-limited rather than distance-limited, it is quite challenging to ascertain a representative physical size for the sources in a given luminosity range. Both the $FEE$ and the size of a source depend on whether we can resolve their emission. The resolution though is a function of the distance, and at the same time more luminous galaxies are seen at larger distances. Hence, upper limits on the sizes of sources also depends on their luminosities. If we simply compute the mean core size of LIRGs using only those systems that are resolved, which are found above the dotted line in Figure~\ref{f:feevslir}b, the result is 3.1\,kpc, while the median is 2.7\,kpc. A more thorough analysis can be performed using the survival analysis package ASURV\footnote{The package ASURV is available at http://astrostatistics.psu.edu/statcodes/asurv}, which properly takes into account the lower and/or upper limits in a given distribution. If we use this approach, we find that the mean core size of LIRGs at 13.2$\,\micron$ is 2.6$\pm$0.1\,kpc.



Regarding the ULIRG sub-sample, establishing a typical size for the region from which their extended emission originates is even more challenging since all of them are unresolved. If we consider the ULIRGs located up to a distance of $\sim\,100\,$Mpc as the reference, we can estimate an upper limit for their core sizes of $\sim\,1.5\,$kpc, which is agreement with previous findings \citep[e.g.,][]{Soifer2001}.

In order to put our results for the GOALS sample in the context of the less luminous but more numerous galaxies of our local universe, we have analyzed the low luminosity sources from the starburst sample of \citep{Brandl2006}. We selected galaxies with $\LIR\,\leq\,10^{11}\,\Lsun\,$ where the IRS slit was well centered on their nuclei (see their Figure~1): IC~342, Mrk~52, NGC~520, NGC~660, NGC~1097, NGC~1222, NGC~3628, NGC~4088, NGC~4676, NGC~4945, NGC~7252 and NGC~7714. Our measurements are plotted as red squares in Figure~\ref{f:feevslir}a and b. We find that the $FEE_{13.2\mu m}$ of these systems ranges between $\sim\,0.4-0.8$ and is within the values shown by GOALS galaxies with similar IR luminosities. This suggests that the extent of the MIR continuum emission in starburst galaxies and in the low luminosity tail of the GOALS LIRGs is comparable.


Finally we should mention that our study provides no evidence that LIRGs with small $FEE_{13.2\mu m}$ have the highest IR luminosities since the most point-like sources do not appear to be systematically the most luminous in the $5-15\,\micron$ range (see also Figure~\ref{f:medianfee}). Instead, as mentioned above, all LIRGs show the same scatter in the $FEE_{13.2\mu m}$ at any given distance. It is only above the threshold of $\LIR\,\sim\,10^{11.8}\,\Lsun$ when we observe a significant reduction of the $FEE_{13.2\mu m}$ to values of $\lesssim\,0.2$.

\subsection{Dependence Of The Extended Emission On The Stage Of Interaction}

We have shown that there is a strong evolution of the compactness of the continuum MIR emission at the transition point from LIRGs to ULIRGs. What is the physical process responsible for this? It is known that most ULIRGs are merging systems with clear signs of interactions (e.g., \citealt{Clements1996}; \citealt{Murphy2001}). Could it be that the merging processes in (U)LIRGs which drive the material of the galaxies to their nuclei causing massive star formation, is the same reason that makes them appear more compact in the MIR?

\begin{figure}
\epsscale{1.15}
\plotone{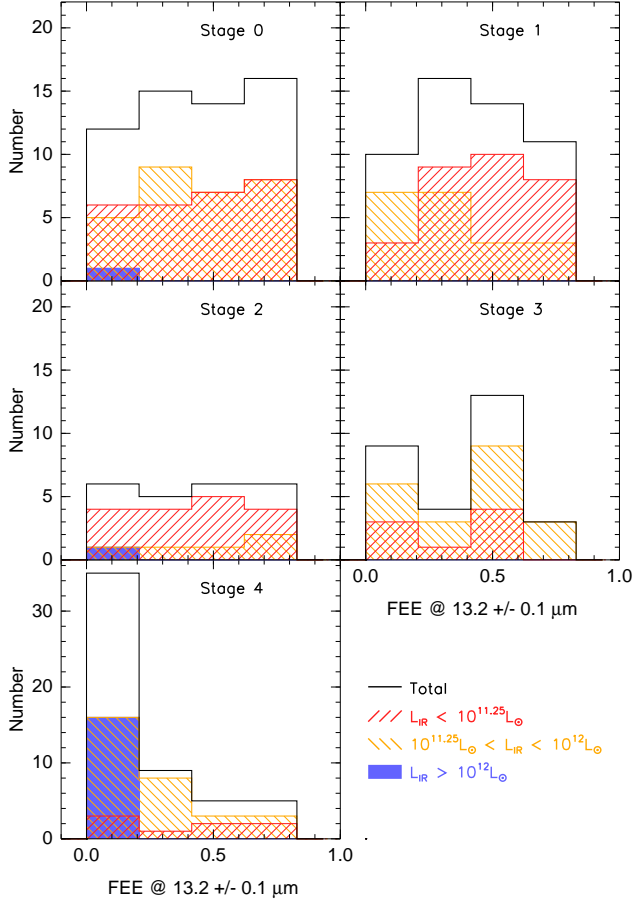}
\vspace{.25cm}
\caption{\footnotesize Histograms of the $FEE_{13.2\mu m}$ for the galaxies in our sample divided as a function of their merging stage as defined by \cite{Petric2010}, from isolated - stage 0 - systems, to fully merged - stage 4 - galaxies \citep[see][for more details]{Petric2010}. The galaxies have been separated in different luminosity bins as in Figure~\ref{f:medianfee}. We exclude from this analysis the galaxies marked with green boxes in Figure~\ref{f:feevslir} (see Section~\ref{ss:feevslir}).}\label{f:feevsmstage}
\vspace{.5cm}
\end{figure}

The galaxies were classified in 5 stages: (0) no obvious sign of disturbance in their IRAC or \textit{HST} morphologies, or published evidence that the gas is in dynamical equilibrium (i.e., undisturbed circular orbits); (1) early stage, where the galaxies are within one arc-minute of each other, but little or no morphological disturbance can be observed; (2) the galaxies exhibit bridges, strong disturbance but they do not have a common envelope and each optical disk is relatively intact; (3) the optical disks are completely destroyed but 2 nuclei can be distinguished; (4) the two interacting nuclei are merged but structure in the disk indicates the source has gone through a merger. The classification was proposed by \cite{Petric2010} and it is based on Hubble Space Telescope (\textit{HST}) imaging, 3.6$\mu m$ \textit{Spitzer} IRAC images as well as optical SDSS images of the objects (see also \citealt{Surace1998PhDT}). In Figure~\ref{f:feevsmstage} we plot total (black) histograms of the $FEE_{13.2\mu m}$ for each interaction stage, and also divided in 3 IR luminosity bins (red, orange and blue, as in Figure~\ref{f:medianfee}). We study 214 (out of the 221) galaxies for which the stage of merger is available. We observe that the most advanced mergers (stage 4) tend to have low $FEE_{13.2\mu m}$ values, i.e., they are more compact. K-S tests performed between the total, stage 4 $FEE_{13.2\mu m}$ distribution and those of the rest of merging stages yield significance values always lower than 0.01. K-S tests performed among the $0-3$ merging stage datasets provide significance values between 0.4 and 0.75. This implies that in fact galaxies classified as to be at merging stage 4 are systematically more compact.

Regarding the IR luminosity bin distributions, we see that merging stage 4 systems are not necessarily the most IR luminous only \citep[see also][]{Rigopoulou1999} but 15\% (8/54) have IR luminosities lower than $10^{11.25}\,\Lsun$. On the other hand, we find that the vast majority, 89\% (16/18), of ULIRGs are systems classified as mergers in their final stage of interaction and all have $FEE_{13.2\mu m}\,\leq\,0.2$, in agreement with previous findings. Similarly, galaxies at the stage 4 of interaction with $10^{11.25}\,\Lsun\,\leq\,\LIR\,<\,10^{12}\,\Lsun$ also tend to show lower $FEE_{13.2\mu m}$ values (like ULIRGs) than less luminous systems. LIRGs and less IR luminous systems are spread uniformly in the $FEE_{13.2\mu m}$ versus \LIR\ parameter space up to the merging stage 2 (the histograms are flat within the uncertainties).

\subsection{Extended Emission And Presence Of An AGN}

Another reason why LIRGs and ULIRGs may appear compact in the MIR is the presence of a dominant AGN. We know that the AGN activity is more prevalent in high IR luminosity systems \citep[e.g.,][2009]{Veilleux1995}, so we would expect that as the AGN emission starts to dominate the IR energy output of a galaxy, it appears progressively less extended.

\begin{figure}
\epsscale{1.15}
\plotone{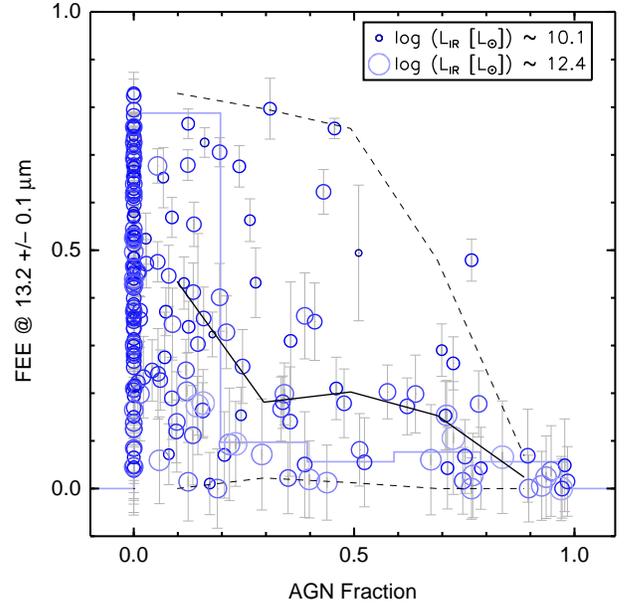}
\vspace{.25cm}
\caption{\footnotesize $FEE$ of the $13.2\micron$ continuum emission as a function of the AGN fraction, estimated by the \PAHd\ equivalent width, for the galaxies of our sample \citep[see][]{Petric2010}. The size of the circles scales with the \LIR\ of the galaxies. The background faint blue line is the normalized histogram of the AGN fraction. The solid black line is the median of the $FEE_{13.2\mu m}$ at the different bins of AGN-fraction. The lower and upper dashed lines are the minimum and maximum $FEE_{13.2\mu m}$ at each bin respectively.}\label{f:feevsagnfrac}
\vspace{.5cm}
\end{figure}

For all galaxies in our sample the contribution of an AGN to their MIR emission has been estimated by \cite{Petric2010} using a number of line and continuum features. We use their estimates of the AGN fraction calculated by means of the so-called ``Laurent diagram'' \citep{Laurent2000}, which is based on the 15$\,\micron$/5.5$\,\micron$ continuum and 6.2$\,\micron$ PAH/5.5$\,\micron$ ratios (see Fig. 3 of \citealt{Petric2010}). By comparing the data in this parameter space with representative ratios of pure AGN, PDR, and \HII\ regions, it is possible to infer the contribution of these components to the total galaxy emission. In Figure~\ref{f:feevsagnfrac} we present the $FEE_{13.2\mu m}$ as a function of this AGN fraction for the 210 galaxies for which it could be derived. The galaxies are again displayed as circles, the size of which scales with their IR luminosity. It is clear that as the AGN fraction approaches unity, the maximum $FEE_{13.2\mu m}$ at the different AGN-fraction bins (upper dashed line) decreases and the galaxies become progressively more compact. In addition, the median $FEE_{13.2\mu m}$ (solid line) also decreases as the AGN fraction increases, although more smoothly. In fact, only 5 out of the 30 AGN-dominated galaxies\footnote{We define as AGN-dominated galaxies those that have an AGN fraction larger than 0.5.} have $FEE_{13.2\mu m}\,>\,0.2$. Independently of the \LIR, it is interesting to see that 60\% (15/25) of the AGN-dominated galaxies with $FEE_{13.2\mu m}\,\leq\,0.2$ are in the final stage of interaction, while only 12\% (3/25) are in the 3rd, 0 in the 2nd, 20\% (5/25) in the 1st and 8\% (2/25) in the 0th (see Figure~\ref{f:feevsmstage}).

We have found that galaxies classified as mergers in their final stage of interaction tend to have low $FEE_{13.2\mu m}$ values and, in particular for ULIRGs, $FEE_{13.2\mu m}\,<\,0.2$. We also show now that the fraction of ULIRGs in merging stage 4 that are AGN-dominated is 60\% (9/15). The fraction of galaxies with $10^{11.25}\,\Lsun\,\leq\,\LIR\,<\,10^{12}\,\Lsun$ and $FEE_{13.2\mu m}\,\leq\,0.2$ (first $FEE_{13.2\mu m}$ bin in Figure~\ref{f:feevsmstage}, stage 4) that are AGN-dominated is 33\% (5/15)\footnote{We have AGN classification for 15 out of 16 galaxies in each of the ULIRG and $10^{11.25}\,\Lsun\,\leq\,\LIR\,<\,10^{12}\,\Lsun$ luminosity bins.} while for lower luminosity systems is 33\% (1/3). Moreover, the fraction of AGN-dominated galaxies classified as mergers at stage 4 with $FEE_{13.2\mu m}\,>\,0.2$ is 0\%, independently of the IR luminosity. That is, \textit{all} AGN-dominated galaxies in the stage 4 of interaction are compact.

\subsection{Extended Emission And Presence Of Cold Dust}


Independently of the process that dominates the IR emission in galaxies (either an AGN or a starburst), we should expect to observe harder radiation fields and higher dust temperatures within more compact environments. In this section we explore whether the extended emission we detect in the $5-15\micron$ wavelength range, that traces warm small dust grains and molecules, depends on the amount of cold dust of our sources. We first calculate the \textit{IRAS} log($f_{60\,\mu m}/f_{100\,\mu m}$) color of the galaxies, which is a well known IR broad-band AGN diagnostic (\citealt{deGrijp1985}), but we find that it does not correlate with the extent of the MIR continuum emission. We also explore whether there is a dependence on the \textit{Spitzer} log($f_{24\,\mu m}/f_{70\,\mu m}$) and log($f_{70\,\mu m}/f_{160\,\mu m}$) colors but do not find any clear trend. However, when we plot in Figure~\ref{f:feevsiras10060} the $FEE_{13.2\mu m}$ as a function of the FIR \textit{IRAS} log($f_{60\,\mu m}/f_{100\,\mu m}$) color, a correlation is visible. Despite the large scatter, galaxies with cold FIR colors appear to be more extended in their MIR continuum. The Spearman rank correlation coefficient is $-0.67$ with a significance level of effectively 0, while the Kendall test provides a value of $-0.49$, implying that the correlation is real. We fitted our data with an outlier-resistant linear fit algorithm provided by the IDL function {\sc robust\_linefit}\footnote{The procedure {\sc robust\_linefit.pro} can be found in http://spectro.princeton.edu/goddard\_doc.html}, and the best fit parameters are given in the following equation:

\begin{figure}
\epsscale{1.15}
\plotone{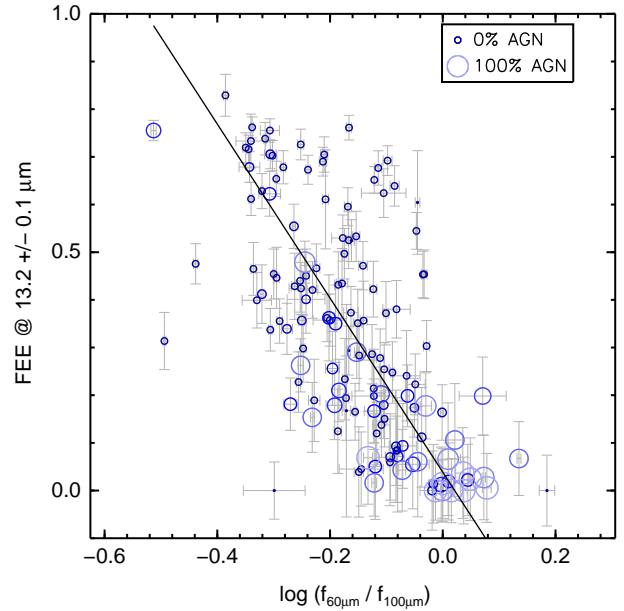}
\vspace{.25cm}
\caption{\footnotesize $FEE$ of the $13.2\micron$ continuum emission as a function of the \textit{IRAS} log($f_{60\,\mu m}/f_{100\,\mu m}$) color. The sizes of the circles scale with the AGN fraction of the galaxies. Points indicate galaxies for which the AGN-fraction was not available. The solid line is the linear fit to the data.}\label{f:feevsiras10060}
\vspace{.5cm}
\end{figure}

\begin{equation}\label{e:feevsfir}
FEE_{13.2\mu m}=0.04\pm0.02-(1.83\pm0.11)\times\log(\frac{f_{60\,\mu m}}{f_{100\,\mu m}})
\end{equation}

\noindent
where $FEE_{13.2\mu m}$ is the fraction of the extended emission of a given galaxy at $13.2\,\micron$, and $f_{60\,\mu m}$ and $f_{100\,\mu m}$ are the \textit{IRAS} flux densities ($f_\nu$) at 60 and $100\,\micron$, respectively. The data have a scatter of 0.2 in the y-axis and 0.1 in the x-axis. The uncertainties in the parameters of the fit have been calculated using bootstrapping resampling analysis. In order to avoid interacting systems for which only one \textit{IRAS} color has been measured due to the limited spatial resolution, we have used only 142 sources for which there is a unique association between one galaxy (one $FEE_{13.2\mu m}$ measurement) and one \textit{IRAS} color. In addition to this correlation, Figure~\ref{f:feevsiras10060} also shows that nearly all galaxies whose MIR emission is dominated by an AGN (i.e., they have a high AGN fraction according to the estimates of \citealt{Petric2010}) are located at the lower right corner of the plot, which means that they are more compact, and have warmer FIR colors. Even if we exclude AGN dominated systems and consider only starburst galaxies with an AGN fraction close to zero, the correlation is still present.

This result has an interesting implication for the forthcoming studies of the cold dust emission of LIRGs/ULIRGs with the \textit{Herschel Space Telescope}. Figure~\ref{f:feevsiras10060} suggests that ``cold'' FIR-selected (U)LIRGs will likely present large $FEE_{13.2\mu m}$ values. As a result one would expect that a large fraction of their MIR continuum emission originates in their extended component
(see previous section). Furthermore, based on the study by \cite{Bothun1992}, the \textit{IRAS} log($f_{60\,\mu m}/f_{100\,\mu m}$) colors of our galaxies suggest that they have very strong dust temperature gradients (their Section~4.1 and Table~3). Consequently the cold, FIR emission from these systems will be more extended than the warmer MIR emission. Taking this into account, the $FEE$ we derive for the MIR continuum emission based on our \textit{Spitzer}/IRS observations places a lower limit to the extent of the FIR continuum emission. Therefore, despite the limitations of our long-slit MIR spectra, one may obtain a rough estimate on how much of the FIR emission is extended in any LIRG/ULIRG.


\subsection{Implications For High-Redshift SMGs}

By virtue of the negative $K$-correction, SMGs are systems with IR luminosities in the ULIRG range, almost independently of their redshift. Recent studies of SMGs, based on various diagnostics including the MIR and FIR wavelength range have suggested that they are systems dominated by star formation (\citealt{Pope2008}; \citealt{Farrah2008}; \citealt{Murphy2009}; \citealt[2008, 2009]{Sajina2007}). In addition, interferometric observations of their CO content and radio continuum emission reveal that they host large amounts of dust and molecular gas ($\MH2\,\sim\,10^{10-11}\,\Msun$) which may fuel prodigious star formation rates of several hundred solar masses per year (\citealt{Neri2003}; \citealt{Chapman2004}; \citealt{Greve2005}; \citealt[b]{Daddi2009a}; \citealt{Tacconi2008}). Analysis of the kinematics of their ionized gas emission using the \Halpha\ emission red-shifted in the near-infrared, indicate that SMGs as well as some high-redshift ULIRGs are extended up to several kpc scales \citep[e.g.,][]{Alexander2010}. In some cases they display motions which can be attributed to organized rotating disks and in others they appear to belong to interacting systems (\citealt{Tacconi2008}; \citealt{Bothwell2010}; \citealt{Ivison2010}). Recent studies of the FIR continuum and [CII]\,158$\,\micron$ emission of high-$z$ IR luminous galaxies also suggest that the star formation in these sources is extended over several kpc (\citealt{Kovacs2010}; \citealt{HD2010}).

It is well accepted that in the local universe the MIR emission traces the star formation regions emitting in \Halpha\ and it correlates with the presence of molecular gas. Based on our analysis we find no evidence that the local ULIRGs of the GOALS sample are extended in the MIR to the sizes suggested by the \Halpha\ or CO measurements of high-z sources. It is true that the \textit{Spitzer} spatial resolution is quite limited, but we have shown that even the closest ULIRGs in our sample have unresolved cores and $FEE_{13.2\mu m}\,\lesssim\,0.2$. Instead it is the LIRG population that displays resolved emission extended over several kpc scales. In addition, it appears that the SMG population comprises not only isolated, disk-like galaxies but also merging systems, which is also in agreement with our findings for the LIRG class.

On the other hand, it is still uncertain whether high-redshift IR luminous sources are simple scaled up analogues of local, less luminous systems in terms of all their physical properties. Integrated MIR spectra of SMGs (\citealt{Pope2008}; \citealt{MD2009}) do display PAH equivalent widths more typical of local LIRGs rather than the weaker values found for local ULIRGs (\citealt{Desai2007}). However, some recent works suggest that the star formation law in SMGs is similar to that of local ULIRGs and is much more efficient than in local, less IR luminous disk-like galaxies, or than in high-$z$ BzK galaxies (\citealt{Daddi2010a}; \citealt{Daddi2010b}; \citealt{Genzel2010}). We will address the issue of the spectral properties of star formation in local LIRGs and ULIRGs in our next paper, where we analyze separately the nuclear and extended emission of the PAH features, emission lines and the 9.7$\,\micron$ silicate feature of the galaxies in the GOALS sample.


\section{Conclusions}\label{s:summary}

We analyzed the spatial profiles of low-resolution 5-15$\mu$m \textit{Spitzer/IRS} spectra of the GOALS sample and quantified the extent of their MIR emission, $FEE_\lambda$. Our work indicates that:

\begin{itemize}

\item There is a diversity in the shape of the $FEE_\lambda$ as a function of wavelength. The variation in the spatial extent of the various MIR features such as PAHs, emission lines, and continuum implies that the MIR emission in (U)LIRGs is complex. However, we find 3 types of $FEE_\lambda$ functions: constant/featureless, PAH-/line-extended, and silicate-extended. Several physical processes, such as AGN emission as well as multiple bursts of star formation, nuclear and extra-nuclear, are suggested to produce the integrated MIR spectrum of (U)LIRGs and to determine their spatial extent at different wavelengths.

\item More than 90\% of the galaxies in the GOALS sample have median $FEE_\lambda$ larger than 0.1. Furthermore, more than 30\% of the galaxies have median $FEE_\lambda$ larger than 0.5, implying that at least half of their MIR emission is extended. As a whole, the median $FEE_\lambda$ of local LIRGs is $\sim\,2-3$ times larger than that of ULIRGs.


\item Despite the limited spatial resolution of the \textit{Spitzer}/IRS spectra, we find a steep decrease in the extent of the continuum emission, $FEE_{13.2\mu m}$, at the threshold of $\LIR\,\sim\,10^{11.8}\,\Lsun$. While LIRGs display a wide range of $FEE_{13.2\mu m}$, spanning from compact objects to sources extended up to 85\%, galaxies above this threshold show unresolved cores and very compact emission, in particular ULIRGs, which all have $FEE_{13.2\mu m}\,\lesssim\,0.2$ independently of their distance. We measure galaxy core sizes (FWHMs) of LIRGs at 13.2$\,\micron$ up to $10\,$kpc, with a mean of 2.6\,kpc if upper limits to the sizes of unresolved galaxies are taken into account. If only resolved sources are considered, the mean core size of LIRGs is 3.1\,kpc. Our estimate for those of ULIRGs is less than $1.5\,$kpc.

\item Galaxies classified as mergers in their final stage of interaction, based on imaging from the \textit{Hubble} and \textit{Spitzer Space Telescopes}, show lower $FEE_{13.2\mu m}$ values than galaxies in earlier stages. Galaxies with $10^{11.25}\,\Lsun\,\leq\,\LIR\,<\,10^{12}\,\Lsun$ in their final stage of interaction also display a similar trend (like ULIRGs), showing lower $FEE_{13.2\mu m}$ values than less luminous systems.

\item The maximum and the median $FEE_{13.2\mu m}$ decrease as the AGN-fraction increases. Galaxies with AGN-fractions larger than 50\% are more compact, and 60\% of those that have $FEE_{13.2\mu m}\,<\,0.2$ are systems in the final stage of interaction. Furthermore, \textit{all} AGN-dominated galaxies classified as mergers in their final stage of interaction have $FEE_{13.2\mu m}\,<\,0.2$, i.e., are compact, independently of their \LIR.

\item The $FEE_{13.2\mu m}$ and the \textit{IRAS} log($f_{60\,\mu m}/f_{100\,\mu m}$) color appear to be correlated, with the MIR continuum emission of colder galaxies being more extended. Due to the large temperature gradients present in our galaxies, the $FEE$ of the MIR continuum provides a rough lower limit to the $FEE$ of the FIR emission in these systems.

\end{itemize}

Determining the extent of a galaxy at MIR, FIR, and even radio wavelengths can yield information about how the stellar populations are distributed in it, and therefore reveal how the galaxy has formed and evolved. These questions will be soon addressed by new observations obtained with the \textit{Herschel Space Telescope} and the Atacama Large Millimeter Array (ALMA), and in the more distant future with the \textit{James Webb Space Telescope}.

\acknowledgments 

TD-S wants to thank E. da Cunha for her suggestions and interesting discussions.
TD-S and VC would like to acknowledge partial support from the EU ToK grant 39965 and FP7-REGPOT 206469. This research has made use of the NASA/IPAC Extragalactic Database (NED), which is operated by the Jet Propulsion Laboratory, California Institute of Technology, under contract with the National Aeronautics and Space Administration, and of NASA's Astrophysics Data System (ADS) abstract service.\\

\appendix \label{a:appendix}

\section{Obtaining the $FEE_\lambda$ Functions}\label{sa:analysis}

\subsection{Detailed Data Reduction and Calculation of the $FEE_\lambda$}\label{ssa:feecalc}

In order to calculate the fraction of extended emission ($FEE_\lambda$) as a function of wavelength, we used as input data the 2 dimensional IRS coa2ds files obtained as explained in \cite{Petric2010}. Due to the limited spatial resolution of \textit{Spitzer}, we performed the analysis using only the SL module (see Section~\ref{ss:obs}), which covers a wavelength range from $\sim\,5.5\,\micron$ to $\sim\,14.5\,\micron$ (SL2 order: $\sim\,5.5\,\micron$ to $\sim\,7.5\,\micron$; SL1 order: $\sim\,7.5\,\micron$ to $\sim\,14.5\,\micron$). Details regarding the steps taken for the reduction and creation of the coa2ds can be found in \cite{Petric2010}. We used version 18.7 of the \textit{wavesamp} file which contains the information on the projection of the spatial and spectral elements of the IRS slits onto the 2 dimensional detector array via a set of pseudo-rectangles. Each pseudo-rectangle defines the spatial direction that crosses an IRS spectrum at a given wavelength. Therefore, to extract the spatial profiles of a galaxy as a function of wavelength, we interpolated the IRS spectrum along the direction traced by each pseudo-rectangle. The same approach was taken to produce the spatial profiles of the standard star used as our reference point spread function (PSF) representing an unresolved source. Each spatial profile was fitted with a Gaussian function leaving as free parameters its maximum intensity, position, and full width half maximum (FWHM). Indeed, the later gives us a measure of the size of the source's ``core'' at each wavelength (i.e., the Gaussian component of the extended emission; see Figure~\ref{f:feevslir}b). To calculate the unresolved emission of the galaxy, we scaled the maximum of the PSF spatial profile to the maximum of the galaxy spatial profile and subtracted it. The remaining flux is the extended emission (EE) of the galaxy. We also measured the total emission of the galaxy by integrating the flux over the entire spatial profile defined by the IRS slit. The EE divided by the total emission of the galaxy is the fraction of EE (FEE). This calculation was repeated for each wavelength resolution element (pseudo-rectangle) and consequently the $FEE$ is a function of $\lambda$ ($FEE_\lambda$). The calculation of the $FEE_\lambda$ has some advantages over other methods used to ascertain the compactness of a galaxy, such as the concentration index or the Gini coefficient. The concentration index relies on aperture photometry and therefore does not provide an estimate the shape of the spatial distribution of the PSF, mixing both unresolved and resolved emission. The Gini coefficient is also totally independent of the shape of the galaxy. As a consequence the growth of the physical size of the unresolved component of the galaxy with distance is not taken into account and it is intrinsically included in the Gini statistics, which would make the interpretation of our results even more challenging."



\begin{figure*}
\epsscale{.38}
\plotone{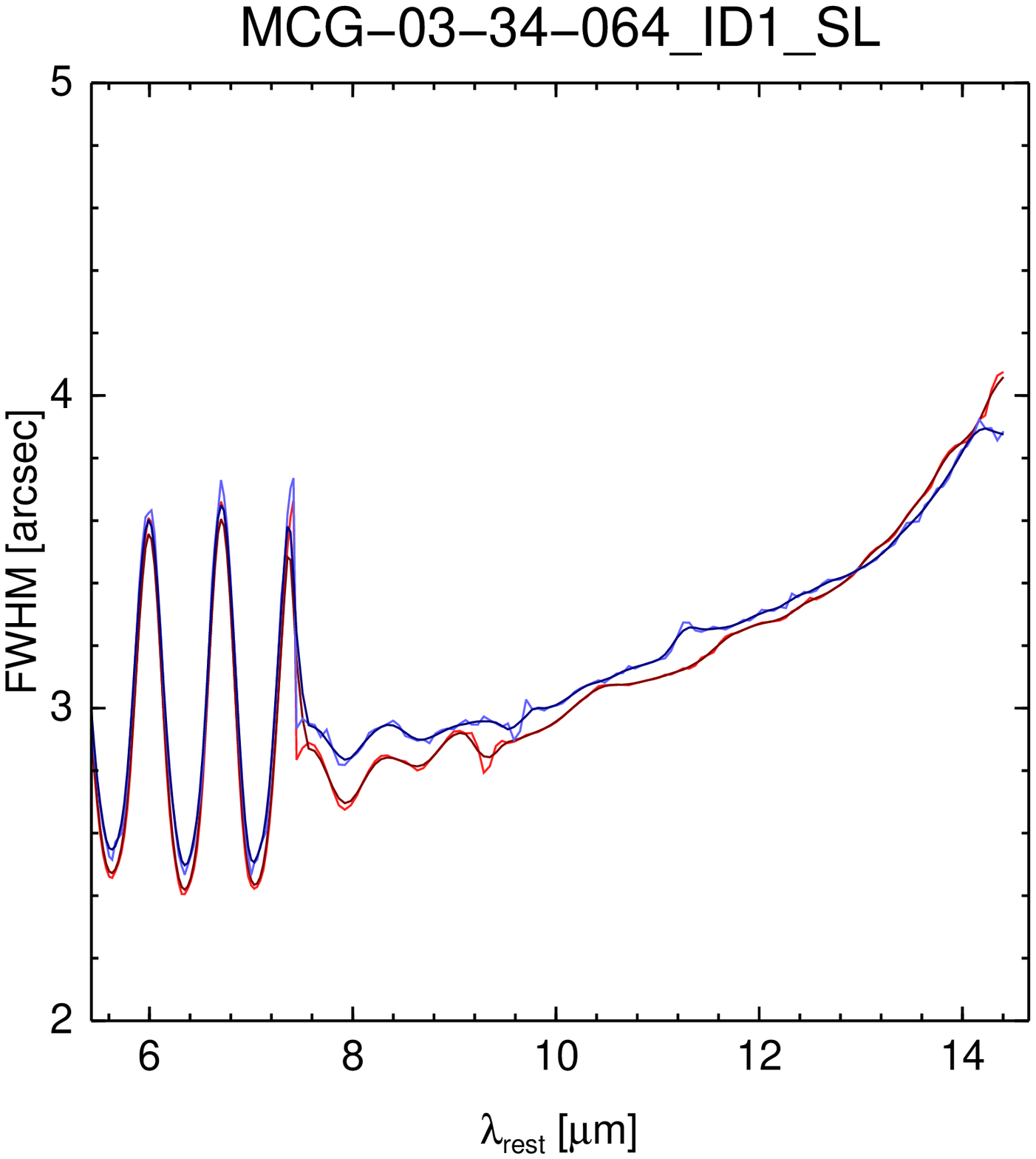}\plotone{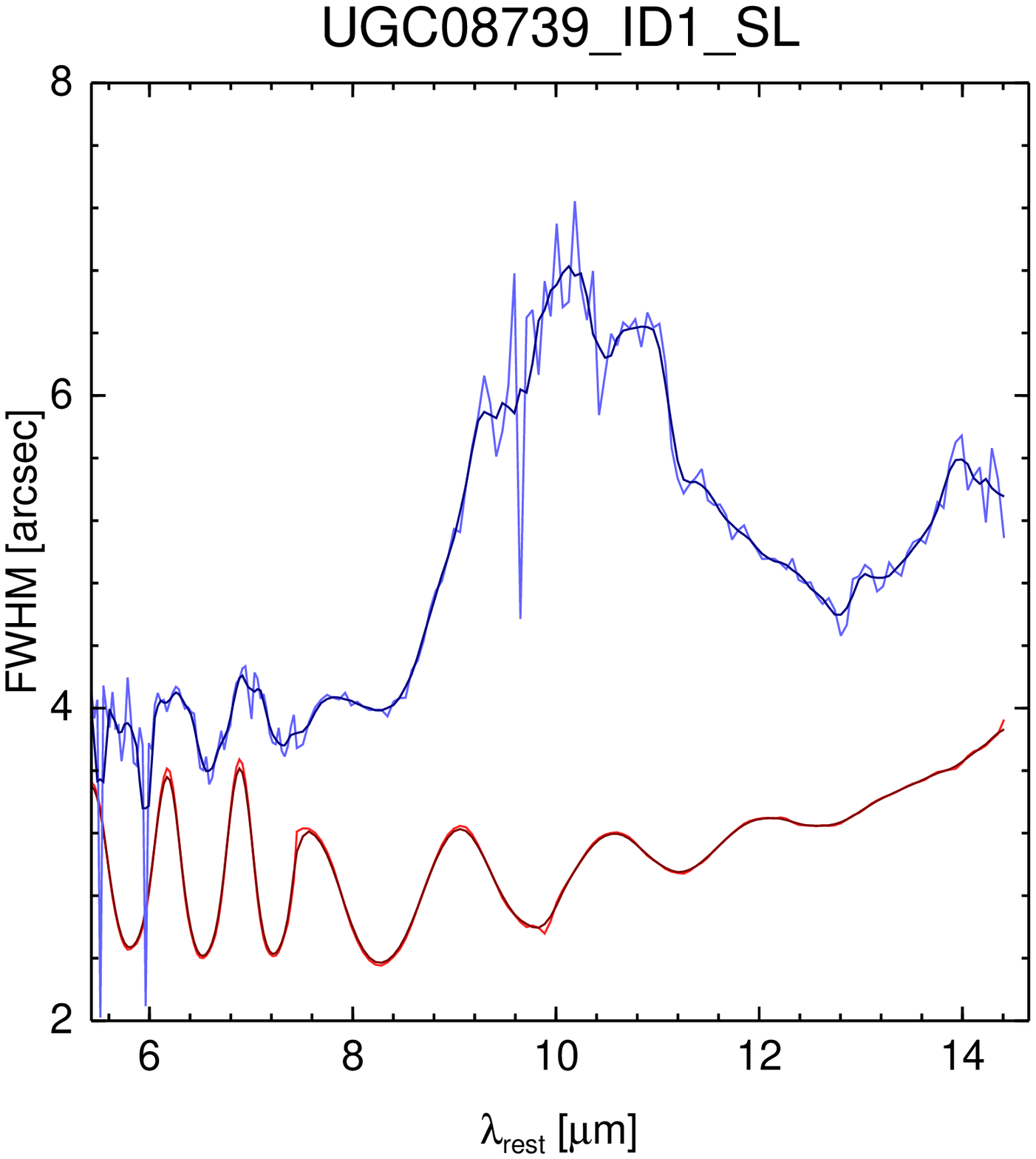}\plotone{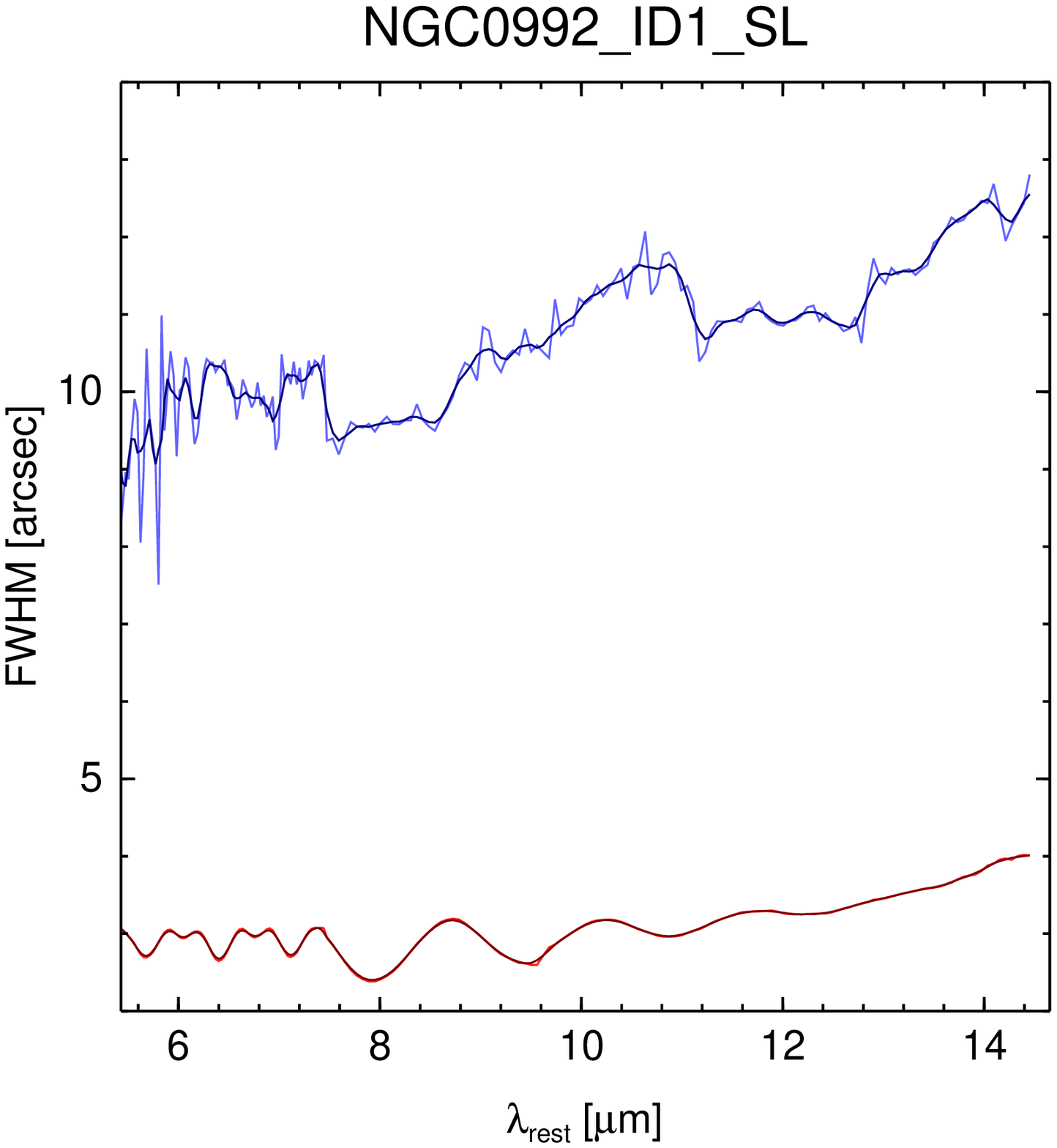}
\plotone{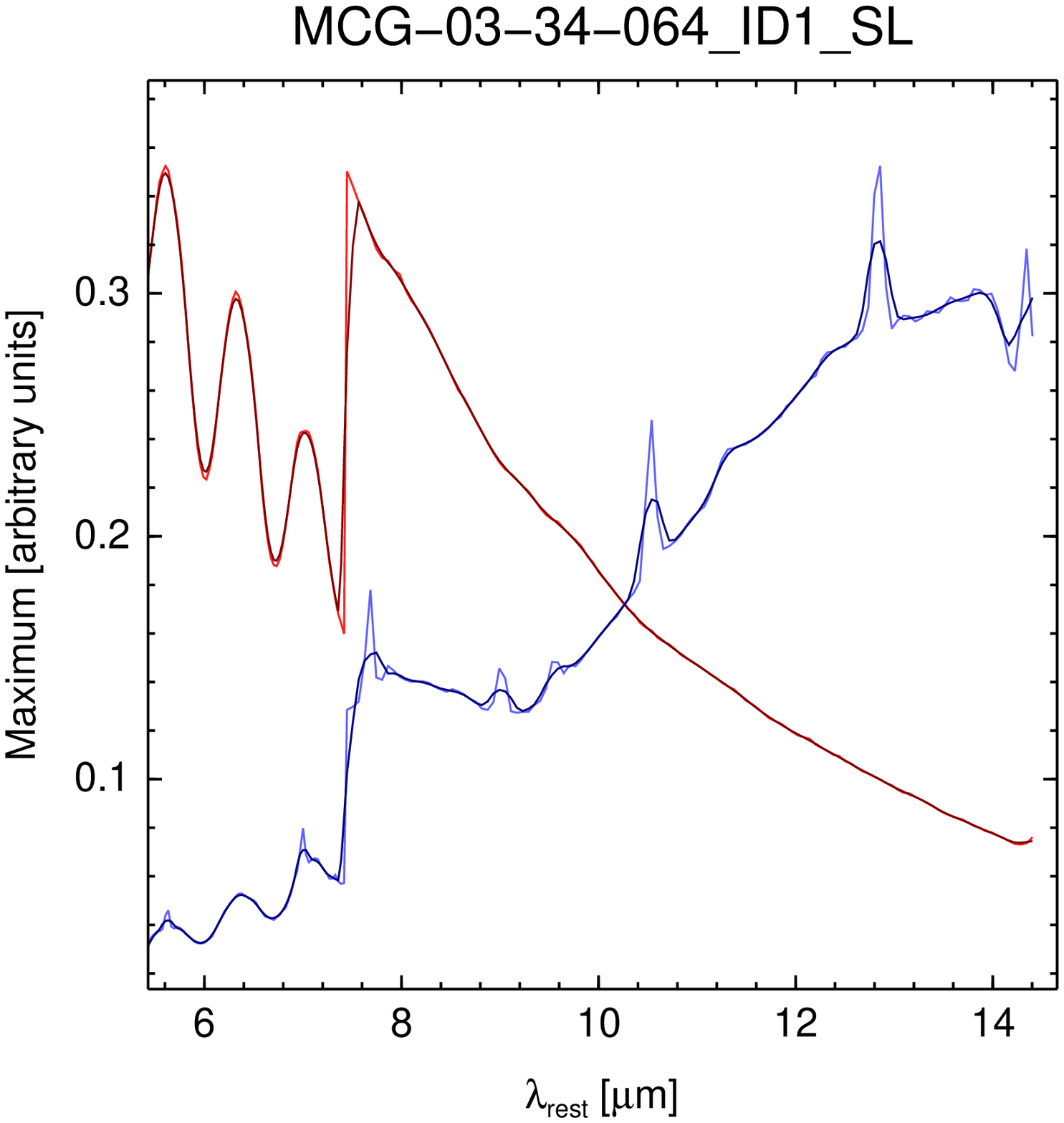}\plotone{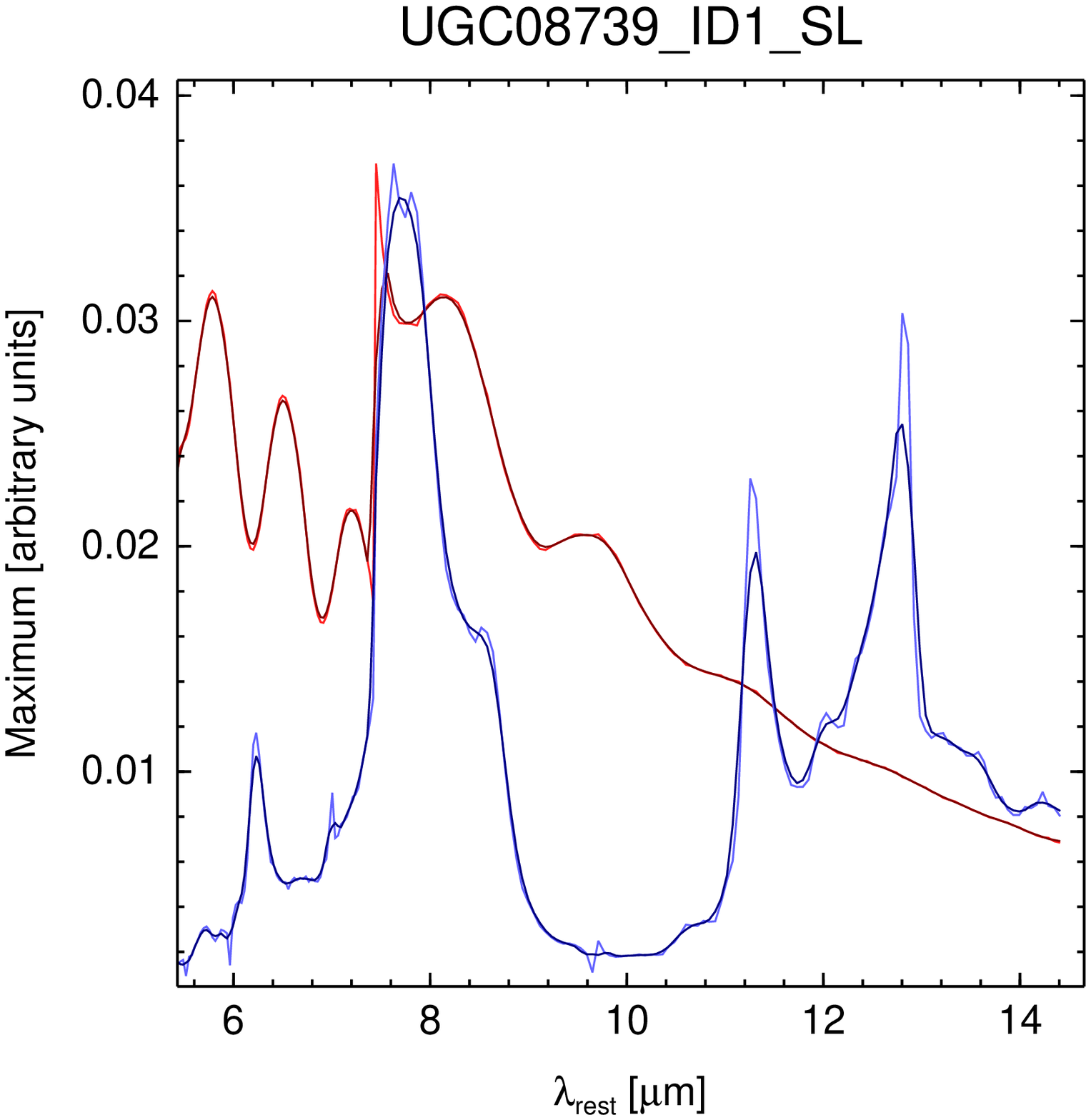}\plotone{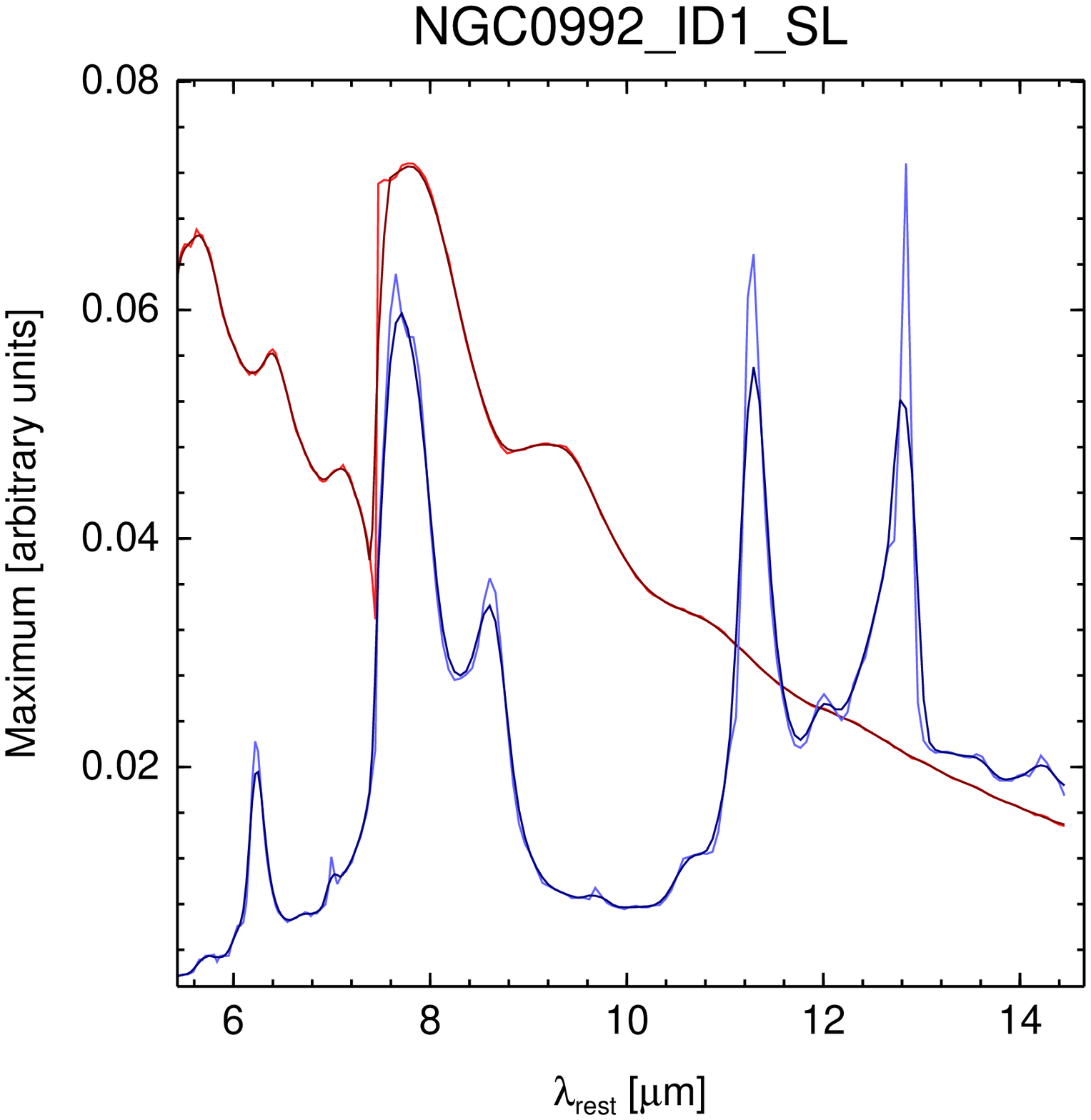}
\plotone{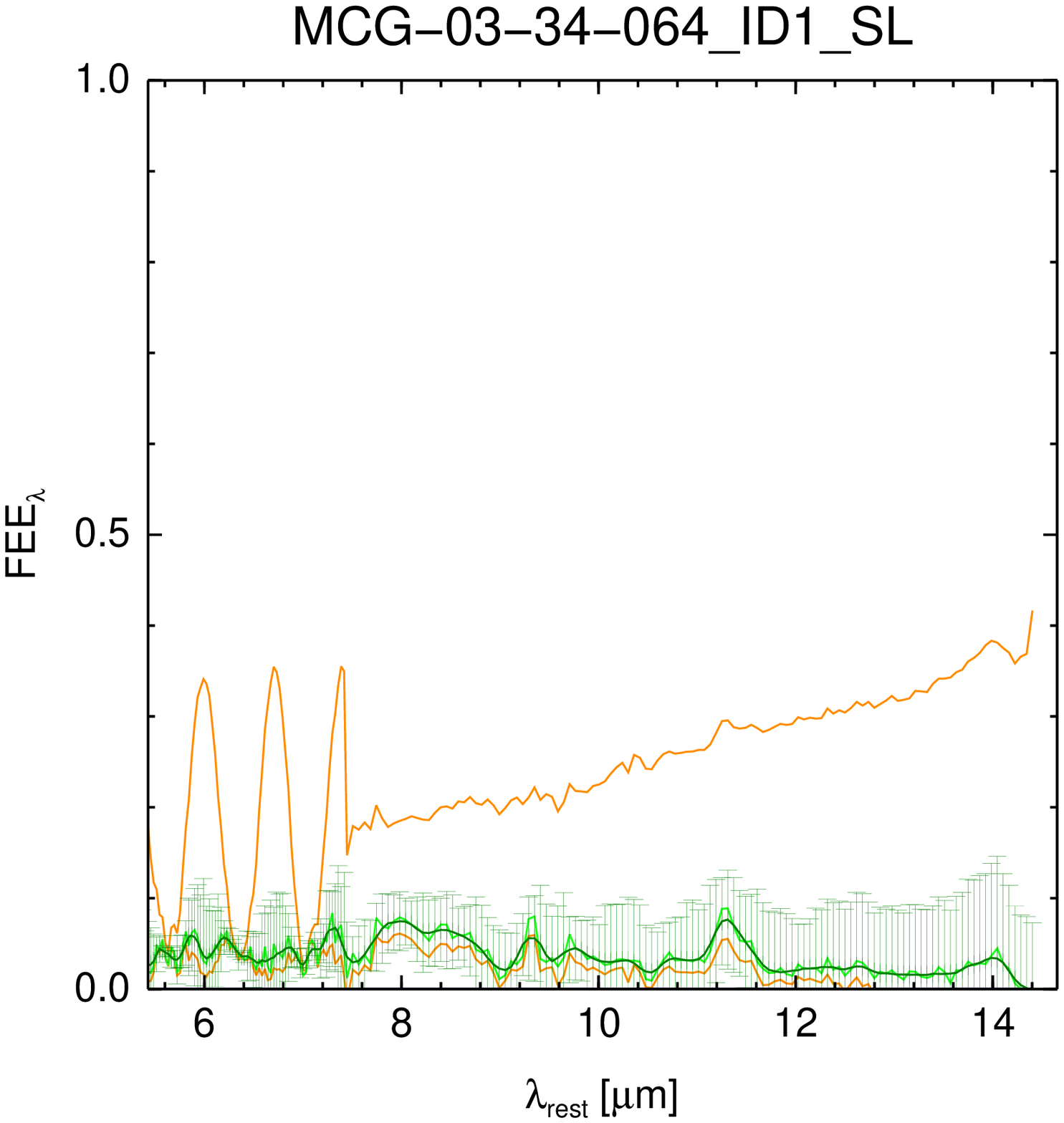}\plotone{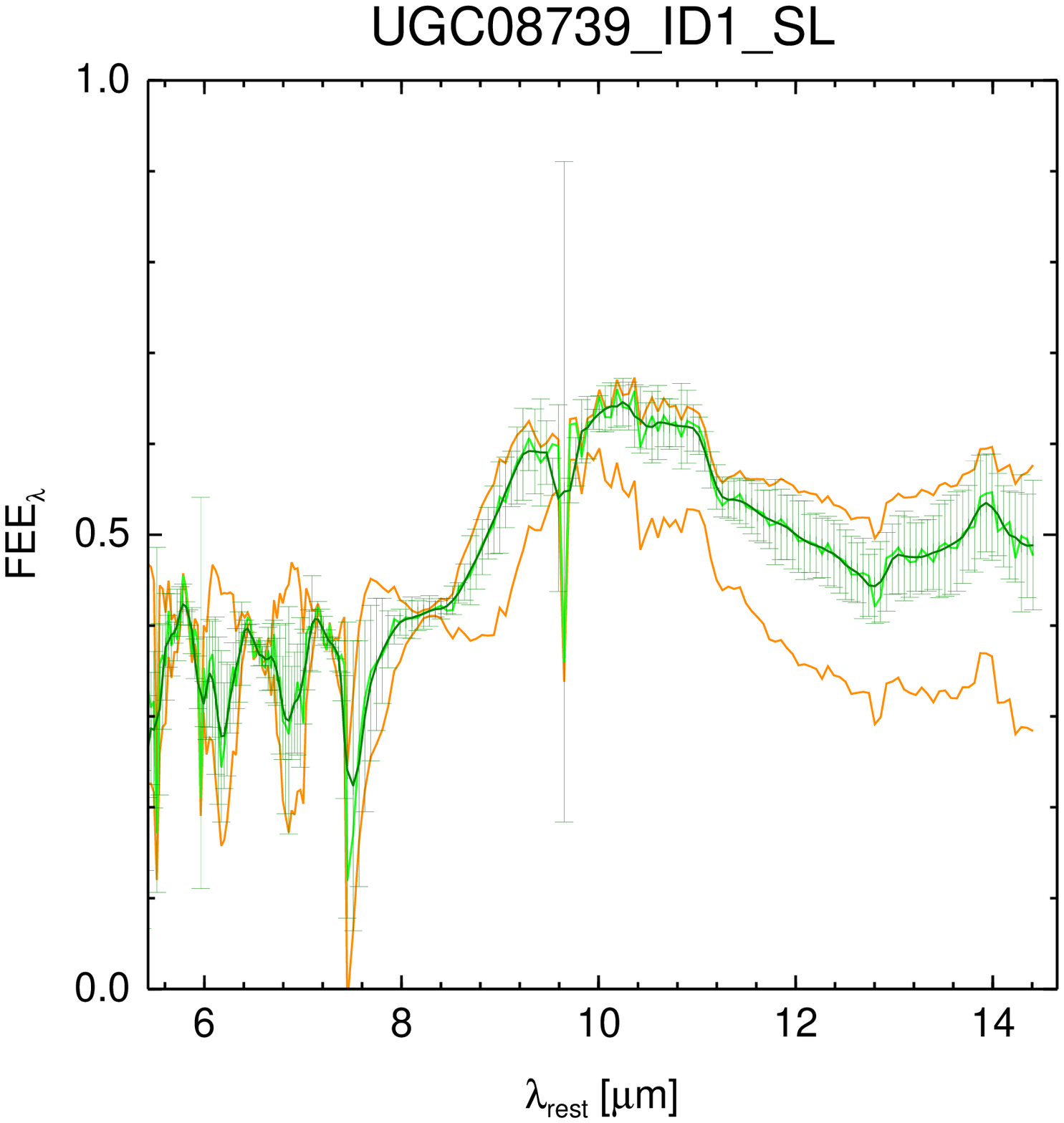}\plotone{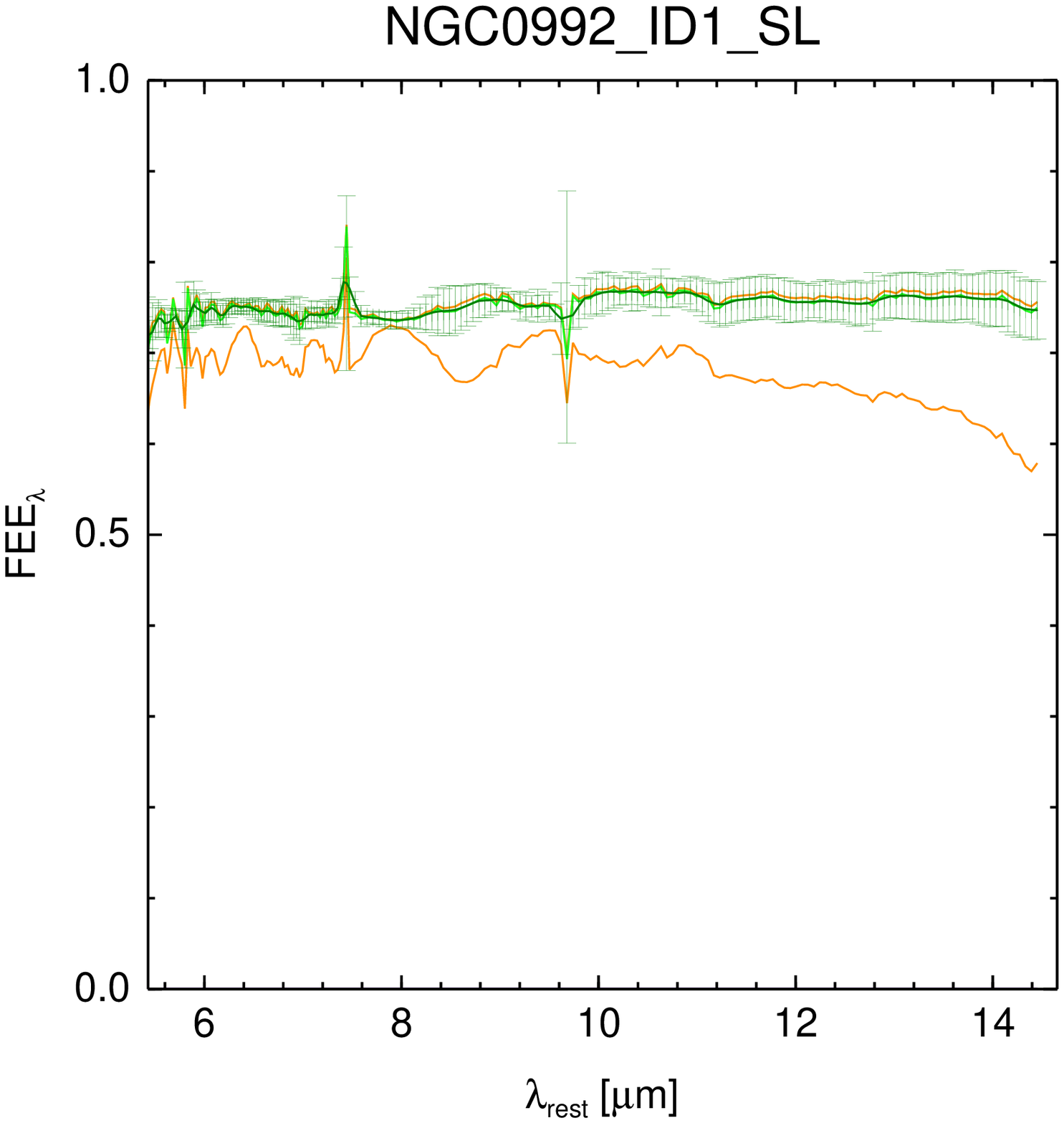}
\vspace{.5cm}
\caption{\footnotesize Top panel: FWHM of the Gaussian function for three galaxies of the sample as a function of wavelength and averaged for the two nod positions (blue lines). The FWHM of the PSF is also shown for reference (red lines). The galaxy on the left is unresolved while the other two show FWHMs larger than those of the PSF. Middle panel: Maximum of the Gaussian function (arbitrary scaled) for the same three galaxies as a function of wavelength and averaged for the two nod positions (blue lines). These plots are equivalent to the spectra of the galaxies, except that they are affected by the under-sampling. The maximum of the PSF (arbitrary scaled at each order) is also plotted for reference (red lines). Bottom panel: Fraction of extended emission, $FEE_\lambda$, for the same three galaxies as function of the wavelength and averaged for the two nod positions (green lines). The bottom orange lines represent the $FEE_\lambda$ that would be obtained if the correction factor was not applied at all, that is, if we considered that the source is totally unresolved. The top orange lines represent the $FEE_\lambda$ that would be obtained if the correction factor was fully applied, that is, if we considered that the source is very extended. See text for details. The thin-colored lines are actual values. The thick-colored lines are the same values but smoothed with a 4-pixel box to reduce the noise.}\label{f:corrfac}
\vspace{.5cm}
\end{figure*}

\subsection{Corrections to the $FEE_\lambda$}\label{ssa:feecorr}

Since the pixel size of the IRS SL detector is $1.8\arcsec\,\times1.8\arcsec$, the IRS spectra are under-sampled below $\sim\,10\,\micron$. The under-sampling affects the spatial distribution of the flux of unresolved sources over the pixels of the detector. As a result, the exact pixel intensity depend on where the source is located on it at a sub-pixel scale. Since a IRS spectrum does not run along the columns of the detector but it displays a slight curvature, the spatial profile of the source at each spectral resolution element is not centered at the same sub-pixel position for all wavelengths. Therefore the under-sampling introduces an intrinsic variation on the shape of the spatial profiles as a function of the wavelength, which is reflected in a change of the FWHM as well as the maximum value of the fitted Gaussian to the spatial profiles. The effect of the under-sampling can be seen in Figure~\ref{f:corrfac} where the FWHM and maximum of the spatial profiles of the PSF are shown as a function of the wavelength (red lines; top and middle panels). We note that the FWHM and maximum may vary as much as 50\% in the SL2 module. This factor is very important because, in order to calculate the amount of extended emission of a galaxy at a given wavelength, we need to scale the maximum of the PSF to the maximum of the profile of our source. If a galaxy is unresolved, its fitted maximum will present the same behavior as the maximum of the PSF and the wiggling pattern created by the under-sampling will cancel out. If, on the contrary, a galaxy is resolved, its spatial profile will not be affected by the under-sampling, but the profile of our PSF will be and will introduce this pattern in the calculation of the $FEE_\lambda$.


Therefore, before the calculation of the $FEE_\lambda$ of a galaxy, we need to correct the maximum of the PSF, in a manner which takes into account the under-sampling in the spatial profile of the galaxy, which in turn depends on its compactness. In order to calculate how compact the galaxy is, we compare its fitted FWHM with that of the unresolved PSF at each wavelength and correct accordingly the maximum of the PSF. A zero correction means that the galaxy is totally unresolved, as is the PSF. A full correction is applied when the galaxy is much more extended than the PSF and implies that we have to correct the maximum of the PSF as if it was not affected by the under-sampling at all. What is this correction factor?

The correction factor to the maximum of the PSF is the value of the FWHM of the PSF normalized to its real value at the same wavelength. This is because even if the FWHM and the maximum of the spatial profile of an unresolved source oscillates as a function of the wavelength due to the under-sampling, the total flux does not. As a consequence, the patterns of the FWHM and the maximum of the PSF are ``synchronized'' (the maximum of one is the minimum of the other; see Figure~\ref{f:corrfac}, top and middle panels, red lines) and thus we can use the value of the FWHM to correct the maximum to its real value. In this we assume that the ``real'' (highest) spatial resolution that can be achieved at a given wavelength is the one defined by the minimum of the lower envelope showed by the pattern of the FWHM function of the PSF.
Values larger than this are affected by the under-sampling and cause the FWHM of the spatial profiles to be over-estimated and, conversely, the maximum to be under-estimated. Therefore, in order to correct for the under-sampling, we multiply the maximum of the PSF by the ratio of its observed-to-minimum (real) FWHM. As we mentioned above, this correction depends on how resolved a galaxy is. In unresolved sources the correction is not necessary, while in very extended galaxies it needs to be fully applied. In order to demonstrate the success and accuracy of the method, in Figure~\ref{f:corrfac} we show the FWHM, maximum and $FEE_\lambda$ function of 3 galaxies resolved in a different degree. The final $FEE_\lambda$ at a given wavelength is the average of the values obtained for each one of the two nod positions of the IRS slits, and the uncertainty is the difference between both.

We can see that the $FEE_\lambda$ function (bottom panel, green line), once corrected for the under-sampling effect, shows almost no residuals of the wiggling pattern that can be seen in the FWHM and maximum functions (top and middle panels) of the galaxies and the PSF. Furthermore, the correction works very well for almost unresolved sources (left panels), very extended sources (right panels), as well as for intermediate-resolved sources (center panels). As an example, we show in the bottom panels, together with the $FEE_\lambda$ function (in green), the $FEE_\lambda$ function that would have been obtained if we had not applied the correction factor at all (lower orange line) or if it had been fully applied (top orange line), that is if we had not taken into account the \textit{a priori} knowledge on how extended our galaxy is. The comparison of the FWHMs of the galaxy and the PSF is therefore essential in order to know to which extent the correction factor must be applied at a given wavelength. This is specially true for galaxies that are slightly resolved at certain wavelengths, as is the case of UGC~08739 (middle panel).

There were a few cases, however, where an accurate $FEE_\lambda$ function could not be accurately recovered for the SL2 order (IC~4734, ESO~279-G011, MCG-02-33-098, CGCG~468-002 and MCG-05-12-006). These are galaxies that show FWHM functions that do not present a wiggling pattern despite they are almost unresolved. Therefore their $FEE_\lambda$ functions from $\sim\,5.5\,\micron$ to $\sim\,7.5\,\micron$ are very uncertain.

\subsection{Additional Tests}\label{ssa:feeverif}

In addition, we checked whether the $FEE$ as a function of the wavelength obtained with the IRS spectra is consistent with the value that can be  calculated from the IRAC images at 8$\,\micron$. To do so we performed a PSF-fitting analysis to the images in the same way as we did for the spatial profiles of the IRS spectra at each wavelength. First, we used a sub-sampled point response function (PRF) for the 8$\,\micron$ IRAC band-pass provided by the SSC and interpolated it to the sub-pixel position in the image where the galaxy was located. Then, we rebined the PRF to the pixel size of the IRAC mosaic, scaled it to the peak of the emission of the galaxy, and subtracted the unresolved emission represented by this scaled-PRF. Finally we measured the remaining flux in an area approximately equal to that used for measuring the integrated emission of the galaxy in the IRS slit (see Section~\ref{s:analysis}). Using Equation~\ref{e:fee} as for the spectroscopy, the $FEE_{8\,\mu m}$ obtained from the IRAC 8$\,\micron$ imaging is the ratio of the remaining emission to the total integrated emission of the galaxy measured in the same aperture.

There are some cases where the $FEE_{8\,\mu m}$ is greater the $FEE$ calculated from the IRS spectra. This is due to the different surfaces used for measuring the emission of the galaxy in each case (for imaging and spectroscopy). Although both apertures have the same area, the former has the shape of a box while the one of the IRS slit is very elongated. Since, as it is said above, the slit was not oriented in a preferred direction over the galaxies, a boxy-shaped aperture used for measuring the emission will generally lead to a higher $FEE$ because of its symmetry around the central region of the galaxy which in many cases is axisymmetric. Therefore, the fact that the $FEE_{8\,\mu m}$ is slightly above than those calculated using the IRS data for some galaxies is totally consistent with the approach taken and shows that both calculations of the FEE, using imaging and spectroscopy, are in agreement.




\end{document}